\documentclass[aps,pra,twocolumn,superscriptaddress]{revtex4-1}

\usepackage{lmodern}
\usepackage{amsmath}
\usepackage{amssymb}
\usepackage{siunitx}
\usepackage{physics}
\usepackage{xfrac}
\usepackage{bm}
\usepackage{graphicx}
\usepackage{xcolor}
\usepackage{hyperref}
\usepackage{ulem}
\usepackage[english]{babel}
\usepackage[utf8x]{inputenc}
\usepackage[T1]{fontenc}
\usepackage{gensymb}
\usepackage{braket}
\usepackage[a4paper,top=3cm,bottom=2cm,left=3cm,right=3cm,marginparwidth=1.75cm]{geometry}

\begin{document}

\title{Reproducibility of high-performance quantum dot single-photon sources}

\author{H\'el\`ene Ollivier}
\affiliation{Centre de Nanosciences et Nanotechnologies (C2N), CNRS, Universit\'e Paris Saclay, 10 Boulevard Thomas Gobert, 91120 Palaiseau, France}
\author{Ilse Maillette de Buy Wenniger}
\affiliation{Centre de Nanosciences et Nanotechnologies (C2N), CNRS, Universit\'e Paris Saclay, 10 Boulevard Thomas Gobert, 91120 Palaiseau, France}
\author{Sarah Thomas}
\affiliation{Centre de Nanosciences et Nanotechnologies (C2N), CNRS, Universit\'e Paris Saclay, 10 Boulevard Thomas Gobert, 91120 Palaiseau, France}
\author{Stephen Wein}
\affiliation{Institute for Quantum Science and Technology and Department of Physics and Astronomy,
University of Calgary, Calgary, Alberta, Canada T2N 1N4}
\author{Guillaume Coppola}
\affiliation{Centre de Nanosciences et Nanotechnologies (C2N), CNRS, Universit\'e Paris Saclay, 10 Boulevard Thomas Gobert, 91120 Palaiseau, France}
\author{Abdelmounaim Harouri}
\affiliation{Centre de Nanosciences et Nanotechnologies (C2N), CNRS, Universit\'e Paris Saclay, 10 Boulevard Thomas Gobert, 91120 Palaiseau, France}
\author{Paul Hilaire}
\affiliation{Centre de Nanosciences et Nanotechnologies (C2N), CNRS, Universit\'e Paris Saclay, 10 Boulevard Thomas Gobert, 91120 Palaiseau, France}
\author{Cl\'ement Millet}
\affiliation{Universit\'e de Paris, Centre de Nanosciences et Nanotechnologies (C2N), F-91120 Palaiseau, France}
\affiliation{Centre de Nanosciences et Nanotechnologies (C2N), CNRS, Universit\'e Paris Saclay, 10 Boulevard Thomas Gobert, 91120 Palaiseau, France}
\author{Aristide Lema\^itre}
\affiliation{Centre de Nanosciences et Nanotechnologies (C2N), CNRS, Universit\'e Paris Saclay, 10 Boulevard Thomas Gobert, 91120 Palaiseau, France}
\author{Isabelle Sagnes}
\affiliation{Centre de Nanosciences et Nanotechnologies (C2N), CNRS, Universit\'e Paris Saclay, 10 Boulevard Thomas Gobert, 91120 Palaiseau, France}
\author{Olivier Krebs}
\affiliation{Centre de Nanosciences et Nanotechnologies (C2N), CNRS, Universit\'e Paris Saclay, 10 Boulevard Thomas Gobert, 91120 Palaiseau, France}
\author{Lo\"ic Lanco}
\affiliation{Universit\'e de Paris, Centre de Nanosciences et Nanotechnologies (C2N), F-91120 Palaiseau, France}
\affiliation{Centre de Nanosciences et Nanotechnologies (C2N), CNRS, Universit\'e Paris Saclay, 10 Boulevard Thomas Gobert, 91120 Palaiseau, France}
\author{Juan Carlos Loredo}
\affiliation{Centre de Nanosciences et Nanotechnologies (C2N), CNRS, Universit\'e Paris Saclay, 10 Boulevard Thomas Gobert, 91120 Palaiseau, France}
\author{Carlos Ant\'on}
\affiliation{Centre de Nanosciences et Nanotechnologies (C2N), CNRS, Universit\'e Paris Saclay, 10 Boulevard Thomas Gobert, 91120 Palaiseau, France}
\author{Niccolo Somaschi}
\email[]{niccolo.somaschi@quandela.com}
\affiliation{Quandela, SAS, 86 rue de Paris, 91400 Orsay, France}

\author{Pascale Senellart}
\email[]{pascale.senellart-mardon@c2n.upsaclay.fr}
\affiliation{Centre de Nanosciences et Nanotechnologies (C2N), CNRS, Universit\'e Paris Saclay, 10 Boulevard Thomas Gobert, 91120 Palaiseau, France}


    \begin{abstract}
Single-photon sources based on semiconductor quantum dots have emerged as an excellent platform for high efficiency quantum light generation. However, scalability remains a challenge since quantum dots generally present inhomogeneous characteristics. Here we benchmark the performance of fifteen deterministically fabricated single-photon sources. They display an average indistinguishability of $90.6 \pm 2.8\% $ with a single-photon purity of $95.4 \pm 1.5\% $ and high homogeneity in operation wavelength and temporal profile. Each source also has state-of-the-art brightness with an average first lens brightness value of $13.6 \pm  4.4 \%$. Whilst the highest brightness is obtained with a charged quantum dot, the highest quantum purity is obtained with neutral ones. We also introduce various techniques to identify the nature of the emitting state. Our study sets the groundwork  for large-scale fabrication of identical sources by identifying the remaining challenges and outlining solutions.
 \end{abstract}

\maketitle

\section{Introduction}
\label{intro}

Quantum light sources are key building blocks for the development of quantum enhanced technologies. For instance, indistinguishable single photons are needed for multipartite quantum cryptography \cite{mattar_device-independent_2018} and for developing quantum networks~ \cite{hensen_loophole-free_2015,Liao2017SatellitetogroundQK}. They also  constitute attractive quantum bits for quantum computing~\cite{taballione_88_2019,wang_multidimensional_2018} that do not suffer from decoherence and can be operated at room temperature~\cite{hosseini_unconditional_2011}. Quantum light is also sought after for quantum sensing applications, be it for sub-shot noise quantum imaging~\cite{brida_experimental_2010}  or superresolution~\cite{forbes_super-resolution_2019}.  

For a long time, single-photon based technologies have relied on heralded single-photon sources, based on probabilistic photon pair generation in non-linear crystals~\cite{mosley_heralded_2008}. However, these sources suffer  from intrinsic limitations, where the multi-photon probability scales linearly with the source brightness and reaching high efficiencies requires difficult multiplexing schemes~\cite{kaneda_time-multiplexed_2015,kaneda_high-efficiency_2018}.

Semiconductor quantum dots (QDs) have emerged as an excellent platform to generate single-photons. After two decades of fundamental investigation and technological developments, a clear path has been laid out to obtain highly efficient sources of high quantum purity single photons~\cite{somaschi_near-optimal_2016,wang_towards_2019}. Quantum dots behave like artifical atoms that can emit single photons on demand. By fabricating a microcavity around a QD, these single photons can be efficiently collected while simultaneously mitigating pure dephasing~\cite{giesz_coherent_2016,loredo_scalable_2016,grange_reducing_2017}.

Despite the excellent performance of QD-based sources, the scalability of the technology remains an open challenge because natural QDs grow at random spatial positions~\cite{yoshie_vacuum_2004} and both the natural and site-controlled QDs show inhomogeneous spectral resonances spanning 2 to 10 meV~\cite{huber_highly_2017,scholl_resonance_2019,juska_towards_2013}. While several groups in the QD community have successfully achieved the fabrication of single-photon sources with state-of-the-art performance~\cite{claudon_highly_2010,zadeh_deterministic_2016,kirsanske_indistinguishable_2017,ding_-demand_2016,liu_high_2018}, most demonstrations still rely on fabrication techniques where the spatial and spectral matching of the QD cavity coupling is not fully controlled. Obtaining a good QD-cavity coupling thus relies on  testing a large number of devices -- sometimes in the thousands. Such a low control in the fabrication process is prohibitive for scalability, since the QD-cavity coupling determines not only the source brightness and its spectral bandwidth through the Purcell effect~\cite{purcell_resonance_1946,auffeves_giant_2007}, but also its degree of indistinguishability~\cite{grange_cavity-funneled_2015,iles-smith_phonon_2017}. More recently, several groups have developed techniques to precisely position the QD in a photonic structure~\cite{davanco_heterogeneous_2017,he_deterministic_2017,schnauber_indistinguishable_2019}. Yet, only a few of them report the performance of more than one or two devices~\cite{liu_solid-state_2019,wang_towards_2019}. How far these technologies are from a large-scale production of identical single-photon sources remains an open question.

\begin{figure} [t]
\centering
\includegraphics[width=0.5\textwidth]{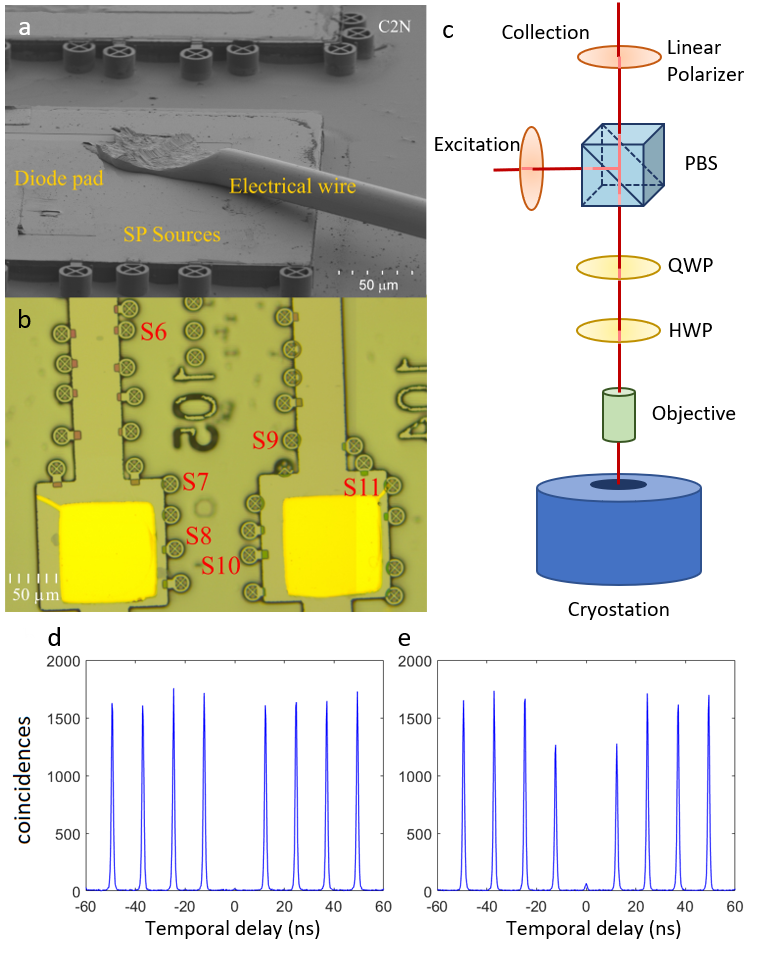}
\caption{\label{fig:fig1} {\bf Single-photon source devices and operating setup.} (a) Scanning electron microscope image of one of the samples under study: each wheel-shaped structure represents a single microcavity coupled to a single QD transition. (b) Optical microscope image of another sample under study. Labels refer to the source numbering used hereafter in the benchmarking. (c) Schematic of the optical setup used for resonant cross-polarization excitation of the QD devices. (d) Typical second-order correlation histogram $g^{(2)}(t)$ as a function of the delay between two detectors and (e) typical second-order correlation in a Hong-Ou-Mandel two-photon interference experiment. }
\end{figure}

In the present work, we address this question by benchmarking fifteen sources consisting of single QDs inside micropillar cavities. The sources are deterministically fabricated using the in-situ cryogenic photolithography technique~\cite{dousse_controlled_2008}. They are operated under resonant excitation in order to obtain the highest degree of quantum purity. For each source, we measure the single-photon purity, indistinguishability, brightness, and study the reproducibility in operation wavelength and temporal profiles.Two types of sources are studied based on different optical transitions: one based on a neutral exciton and the other on a trion, which is a charged exciton. We demonstrate new techniques for identifying these optical transitions, and discuss the physics that determines the source characteristics. Finally, we outline the remaining challenges for larger scale fabrication of identical sources.

\vspace{0,5cm}

\section{Devices and operation}
\label{devices}

A scanning electron microscope (SEM) image and an optical microscope image of two samples are presented in fig. \ref{fig:fig1}a and b respectively. 
The sources that we study are composed of a single semiconductor  InGaAs QD embedded in a micropillar $\lambda$-cavity with 14 (28) GaAs/AlAs Bragg pairs in the top (bottom) mirror. A $20$ nm-thick Ga$_{0.1}$Al$_{0.9}$As barrier, positioned $10$ nm above the QD layer, is used to increase the hole capture time inside the QD \cite{ardelt_controlled_2015}. By creating an electron-hole pair in the QD with an additional laser, we can trap a hole since the electron quickly escapes whereas the hole can not \cite{hilaire_2019}. This technique allows us to optically control the QD charge state. The fabrication process makes use of the cryogenic in-situ lithography technique that allows to position the pillar center within 50 nm of the QD and to adjust the pillar cavity diameter to  ensure the spectral resonance between the QD  and the cavity lines~\cite{dousse_controlled_2008}.   The pillar is connected  through bridges to a circular frame and a large mesa structure where electrical contacts are attached~\cite{nowak_deterministic_2014}.  This allows us to fine-tune the QD-cavity resonance via the Stark effect~\cite{bennett_giant_2010,nowak_deterministic_2014}.\\

In the following, we report on sources from five samples (labelled from A to E) fabricated  from  the same 2-inch wafer grown by molecular beam epitaxy (MBE). Each sample contains 15 to 30 sources. Only a few sources were investigated on samples A through C and all sources were investigated on samples D and E.  Among the studied sources, we selected those giving a first lens brightness greater than $5 \% $. For samples D and E, it was the case for 6 and 4 sources respectively, corresponding to 20 to 25\% of sources per sample. The labels in fig. \ref{fig:fig1}b show the corresponding sources  for sample D.

Highly indistinguishable photons are obtained from QDs when using resonant excitation~\cite{monniello_indistinguishable_2014,he_-demand_2013}. A typical setup for this purpose is presented in fig. \ref{fig:fig1}c. A laser providing 15~ps pulses at a repetition rate of 81 MHz  is set to be resonant with the QD transition.   The excitation beam enters the setup through a single-mode fiber; then a first telescope (not shown in the figure) optimizes the spatial overlap of the beam with the fundamental mode of the micropillar~\cite{dousse_controlled_2008,hilaire_accurate_2018}. An input polarizer sets the polarization along  the reflection axis of a polarizing beam splitter (PBS). The beam is sent through a 0.45 NA objective to a low vibration cryostat where the sample is cooled down to 7~K. The QD emission is collected in the transmission mode of the PBS, orthogonal to the input polarization. A half-wave plate (HWP) and a quarter-wave plate (QWP) are used together to control the polarization with respect to the cavity axes and to correct for polarization ellipticities induced by the setup. In the collection mode of the PBS, a second polarizer improves laser extinction and a second telescope adjusts the single-photon beam diameter to match the collection fiber mode.\\

Fig. ~\ref{fig:fig1}d and \ref{fig:fig1}e present typical second-order correlation histograms characterizing the single-photon purity $1{-}g^{(2)}(0)$ and the photon indistinguishability, here measured on device S7. The $g^{(2)}(0)$ is given by the integrated coincidences around zero delay normalized by the area of the side-band peaks. The indistinguishability is measured via two-photon Hong-Ou-Mandel (HOM) interference in a path-unbalanced Mach-Zehnder interferometer, where two consecutive single photons initially separated by about 12 ns interfere at a beam splitter \cite{loredo_scalable_2016,somaschi_near-optimal_2016}. The raw indistinguishability is given by $V_\text{HOM}= 1 - 2 A_0$ where $A_0$ is the ratio of coincidences at zero delay to the averaged coincidences of the uncorrelated peaks. The examples in fig. ~\ref{fig:fig1} correspond to $g^{(2)}(0)=0.0237 \pm 0.0004$ and $V_\text{HOM}=0.895 \pm 0.002 $.

\section{Two types of sources}
 \label{two_types}
 \begin{figure*}

\includegraphics[width=0.75\textwidth]{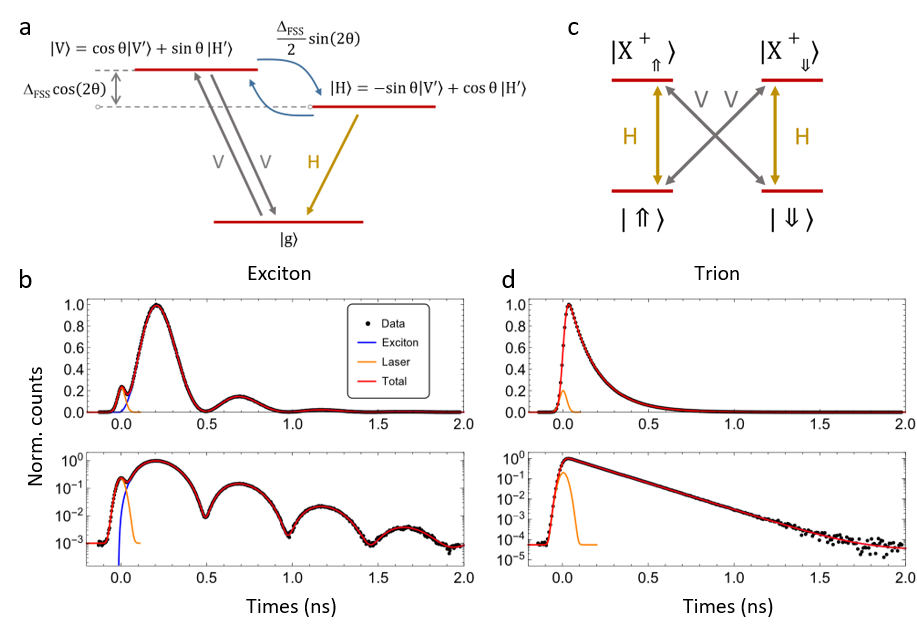}
\caption{\label{fig:fig2} {\bf Two types of sources under study.} (a) and (c) Schematics of the energy levels and optical selection rules for a single-photon source based on an exciton (a) and a trion (charged exciton) (c). (b) Time evolution of the emission for a source based on an exciton (S7), plotted linearly in the upper panel and logarithmically in the lower one. (d) Time evolution of the emission for a source based on a trion (S11). The black points are the experimental data, the red curves are the fits to the theoretical models for total intensity, which include the contribution from the laser (orange curves). The blue curve is the emission arising from the exciton (eq. (\ref{eq_Hproj})) after taking into account the timing jitter of the detector.}
\end{figure*}

Using the cross-polarization setup depicted in fig. \ref{fig:fig1}c, the source operation relies on the ability of the QD optical transition to generate light in a polarization orthogonal to the excitation. The emission process differs substantially depending on whether the optical transition under consideration is an exciton or a trion, which, as we explain in this section, leads to significant differences in the source performance.

\subsection{Exciton based sources}
\label{neutral_exciton_sources}

We first analyze the exciton, which corresponds to a three-level system, as depicted in fig. \ref{fig:fig2}a. It comprises a single ground state $\ket{g}$ and the two intrinsic exciton eigenstates---hereafter labelled $\ket{V'}$ and $\ket{H'}$--- leading to an emission in the corresponding linear polarization $V'$ and $H'$. The two exciton states present a fine-structure splitting (FSS) $\Delta_\text{FSS}= E_{V'} - E_{H'}$ which is an energy difference arising from the Coulomb exchange interaction in a spatially anisotropic QD~\cite{bayer_fine_2002}. 

\begin{figure}[t]
\centering
\includegraphics[width=0.5\textwidth]{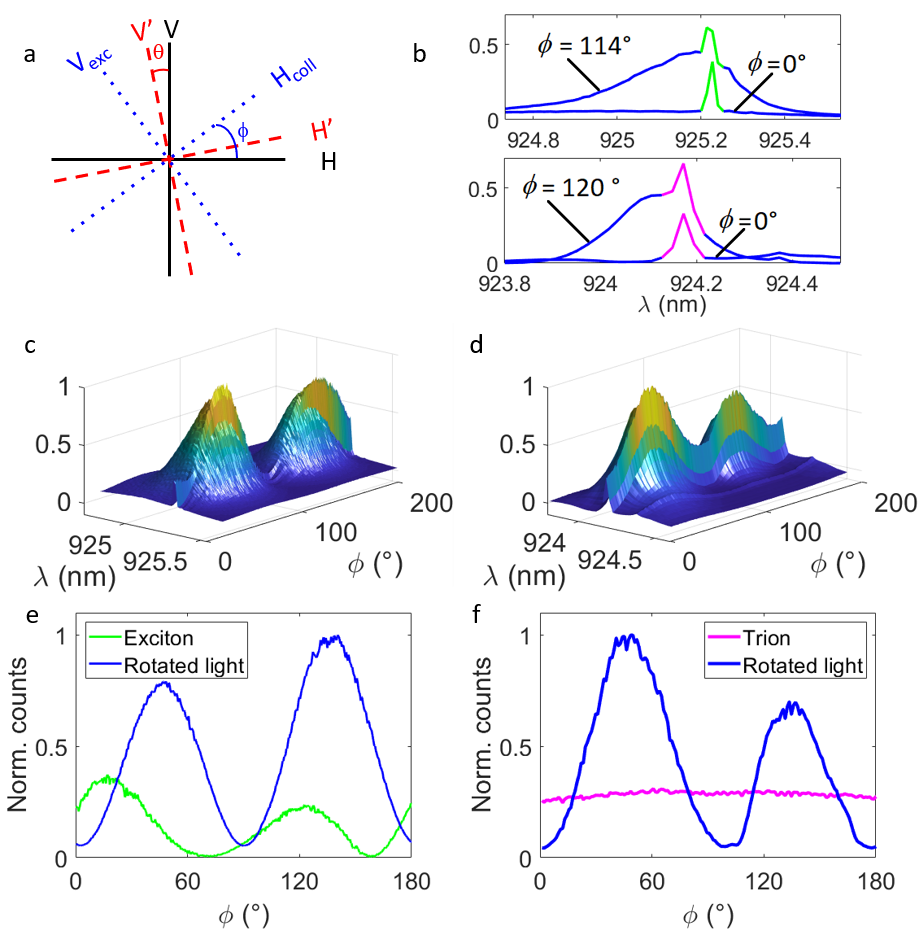}
\caption{\label{fig:fig3} {\bf Line identification through cavity-induced birefringence.} (a) Schematic of the various polarization directions involved in the system: cavity axes $V$ and $H$, exciton axes $V'$ and $H'$ and polarization of the excitation and collection, $V_\text{exc}$ and $H_\text{coll}$ respectively. (b) Spectra obtained for different values of the polarization angle of the excitation laser $\phi$, illustrating the cavity birefringence contribution. The upper panel corresponds to S5, with the narrower signal from the QD highlighted in green, and the lower panel corresponds to S13, with the narrower signal from the QD highlighted in pink. In both panels, the broader blue curve is the cavity rotated light. (c,d) Emission spectra measured as a function of $\phi$ for two devices S5 (c) and S13 (d). The broader emission lines correspond to the signal arising from the cavity birefringence, the narrower lines to the QD emission.  (e) and (f) Intensity of the line arising from the cavity birefringence in blue and from the QD emission line in green (red) as a function of the  angle $\phi$, for the exciton (trion). The cavity rotated light curve in panels (e,f) are the maximum of the broader peak in panels (c) and (d) respectively. The exciton and trion curves are the maximum of the narrower peaks,  minus the linearly interpolated value of the broader peak (cavity rotated light).  }
\end{figure}

The micropillar is not perfectly circular and therefore the cavity itself presents a small anisotropy leading to two nearly-degenerate linearly polarized fundamental cavity modes. The energy difference typically amounts to 30-70~$\mu$eV which is smaller than the cavity spectral linewidth of 150-300~$\mu$eV. These two modes, which we label $V$ and $H$, cause an effective bi-refringence:  when the excitation polarization is not aligned to $H$ nor $V$, the portion of the excitation beam that is coupled to the cavity sees its polarization rotated. Such polarization rotation, recently used to measure the cavity coupling accurately~\cite{hilaire_accurate_2018}, leads to laser signal in the collection path. By choosing the excitation to be parallel with an axis of the cavity, this cavity rotated light is suppressed in the collection mode. In the following, we assume the excitation polarization is parallel to $V$ and the single-photon collection mode is $H$. The quantum states of the exciton that have a dipole parallel to $V$ and $H$ are given by 
\begin{equation}
\left\{
  \begin{array}{rcrcl}
    \ket{V} & = & \cos (\theta) \ket{V'} &+& \sin(\theta) \ket{H'}  \\
    \ket{H} & = & - \sin (\theta) \ket{V'} &+& \cos(\theta) \ket{H'}, \\
  \end{array}
\right.
\end{equation} 
where $\theta$, presented on fig. \ref{fig:fig3}a, is the angle between the optical polarization axis of $V$ and the dipole orientation of the exciton eigenstate $\ket{V'}$.

In a simplified framework where the Purcell effect dominates and pure dephasing effects are negligible, the cavity amplitudes can be adiabatically eliminated to obtain an effective non-hermitian Hamiltonian that describes the emission process~\cite{bertlmann_open-quantum-system_2006}:
\begin{equation}
\begin{aligned}
    H_\text{eff}&=\left (E_{V'}{-}\frac{i\hbar}{2\tau} \right ) \ket{V'} \bra{V'} \\ 
    &{+} \left ( E_{H'}{-}\frac{i\hbar}{2\tau} \right ) \ket{H'} \bra{H'},
\end{aligned}   
\end{equation}
where $\tau$ is the total Purcell-enhanced emission lifetime, which for small cavity asymmetry is assumed to be the same for both excitonic states.

Starting from the excited state $\ket{\psi(t=0)} {=} \ket{V} $ and following a pi-pulse excitation, the time evolution of the exciton is described by
\begin{equation}
\begin{aligned}
\label{eq_X0evol}
\ket{\psi (t)}  & = \cos (\theta) e^{-i \frac{E_{V'} t}{\hbar}} e^{-\frac{t}{2\tau}} \ket{V'} \\
& + \sin(\theta) e^{-i \frac{E_{H'} t}{\hbar}} e^{-\frac{t}{2\tau}} \ket{H'}.
\end{aligned}
\end{equation}

The single-photon emission collected from the $H$ mode is then  proportional to
\begin{equation}
\begin{aligned}
\label{eq_Hproj}
| \braket{H | \psi (t)} |^2  =  e^{-\frac{t}{\tau}}\sin^2\left(\frac{t\Delta_\text{FSS}}{2\hbar}\right)\sin^2(2\theta).
\end{aligned}
\end{equation}

This simple model explains important features. A $V$-polarized excitation creates an exciton state that has no overlap with the $H$ mode at $t=0$. The single-photon emission along $H$ is thus time delayed from the excitation, with a timescale inversely proportional to the FSS. While both components of the excited state decay with the total decay rate $1/\tau$, the emission in the $H$ polarization is modulated by the time-dependent oscillation induced by $\Delta_\text{FSS}$.  In the limit where $\Delta_\text{FSS}\rightarrow 0$, no emission from an exciton in cross-polarized collection mode is expected. In addition, note that if the exciton axes are aligned along the cavity axes so that $\theta=0$, no cross-polarized emission takes place either. On the other hand, the collection is most efficient when $\theta=\pi/4$.

The measured decay dynamics of an exciton is shown in  fig. ~\ref{fig:fig2}b for source S7.  For visualization purposes, the light of the laser was not perfectly extinguished for these data so that we detect both the laser pulse and the exciton emission dynamics to compare the relative timescales. The maximum of the single-photon emission is delayed by approximately $ 200 $ ps from the laser excitation pulse. The overall exponential decay is governed by the Purcell-enhanced spontaneous emission rate, and is also modulated by the phase dependence of the frequency components $H'$ and $V'$ at the rate $\Delta_\text{FSS}$. The experimental observations are accurately reproduced by eq. (\ref{eq_Hproj}) using the following parameters: $\tau = 252{\pm}3$ ps and $\Delta_\text{FSS}=8.58{\pm}0.03$ $\mu$eV. The damping of the oscillations observed in the lifetime curve were found to be consistent with a finite gaussian detector jitter time with a full width at half maximum (FWHM) of $53$ ps, which also dominated the observed width of the 15 ps Gaussian excitation pulse.

All the different excitons studied in this work show a FSS value ranging roughly from $5$ to 10 $\mu eV$. Such  emission process, governed  by the QD and cavity asymmetries, adds extra challenges to the fabrication of identical sources. 
These complex features are circumvented when one considers sources based on a trion, as discussed in the following.

\subsection{Trion based sources}
 \label{trion_sources}

When the QD contains a single charge, here a hole, the energy levels and optical selection rules are different. In the absence of an in-plane magnetic field, no linear polarization direction is favored and the optical transition rules can be written in the $H{-}V$ cavity axes basis. The system ground state is composed of two degenerate energy levels of the hole spin state $\ket{\Uparrow}$  and $\ket{\Downarrow}$, which we define with a quantization axis parallel to $H$. The excited trion states $\ket{X^+_\Uparrow}$ and  $\ket{X^+_\Downarrow}$ correspond to two holes with opposite spin states and one electron. The optical selection rules governing these transitions are summarized in fig. ~\ref{fig:fig2}c: each excited state is connected to both ground states through two optical transitions with linear orthogonal polarizations. 

From this picture, it is straightforward to note that in the absence of spin initialization, namely when the hole spin is a mixture of spin up and down states, a $V$-polarized optical excitation leads to a population of both trion states, that each radiate with $50 \%$ probability along $H$. This happens on the timescale of the excitation pulse, as opposed to the previous case of the exciton.

The illustrated selection rules lead to the generation of single-photon wavepackets with a mono-exponential decay, where the rise time is governed by the excitation pulse length and the decay time by the Purcell-enhanced spontaneous emission rate. Fig. ~\ref{fig:fig2}d shows the emission dynamics of the trion-based single-photon source S11. The emission intensity now shows a rapid rise time followed by a mono-exponential decay with a lifetime of $164.9 \pm 0.9 $~ps. Here, the finite detector response of roughly $50$~ps is consistent with the observed rise time.

As shown in the following, the nature of the optical transition does not only control the temporal profile but it also determines the brightness and single-photon purity of the sources. It is therefore important to develop tools to  identify the nature of the transition. 

Advanced equipment is required to observe the exciton splitting, either for high resolution spectral analysis or, as shown in fig \ref{fig:fig2}b, high temporal resolution of the emission dynamics. In the next section, we demonstrate another simple identification tool based on the polarization-dependent optical selection rules. 

\subsection{Rotated emission identification method}

 \begin{figure*}[t]
\centering
\includegraphics[width=0.9\textwidth]{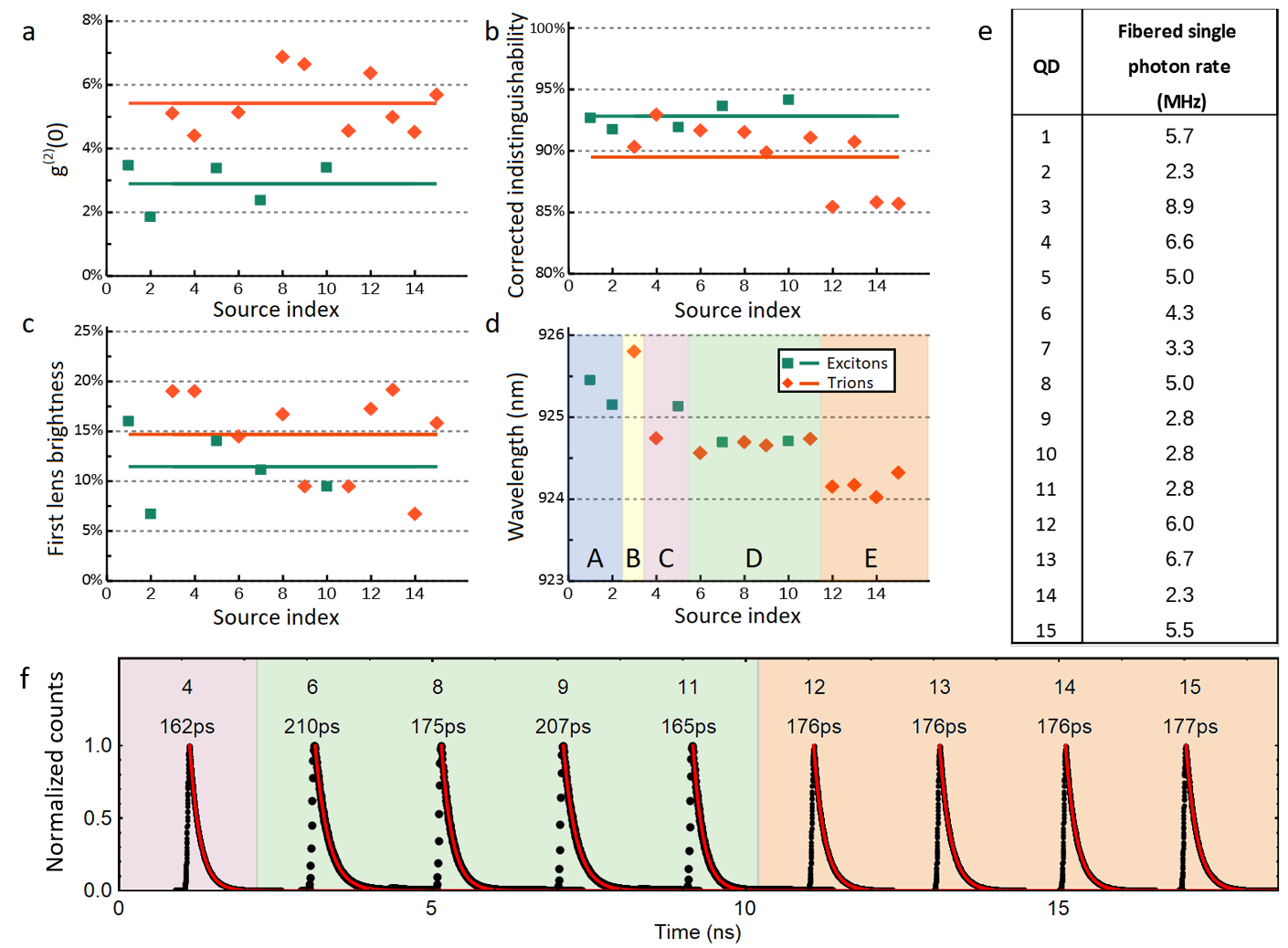}
\caption{\label{fig:fig4} {\bf Performance of fifteen single-photon sources.} (a) Second-order correlation values $g^{(2)}(0)$, characterizing the single-photon purity, (b) mean wavepacket overlap, (c) first lens brightness and (d) operation wavelength. The green squares represent sources based on excitons and the orange diamonds represent sources based on trions. The horizontal solid lines show the mean value for each type of sources. The excitons present an average $g^{(2)}(0)$ of $2.89 \pm  0.74 \%$, an average indistinguishability of $92.8 \pm 1.1 \%$ and an average first lens brightness of $11.5 \pm 3.7 \%$. The trions present an average  $g^{(2)}(0)$ of $5.42 \pm  0.92 \%$, an average indistinguishability of $89.5 \pm 2.8 \%$ and an average first lens brightness of $14.7 \pm 4.6 \%$. (e) Table gathering the measured single-photon rates using a $\sim 30\%$ efficient single-photon detector and their corresponding fibered single-photon rates at the output of a single-mode fiber. (f) Temporal profiles of trion-based sources showing an average exponential decay lifetime of 180 ps with a standard deviation of 17 ps.}
\end{figure*}

The performance of each source is determined both by the nature of the transition and, in the case of excitons, the orientation of the cavity axes with respect to the QD dipoles. In this context, it is useful to analyze the emission collected in cross-polarization when turning the excitation polarization $V_\text{exc}$ by an angle $\phi$ with respect to the cavity axes, see fig. \ref{fig:fig3}a. In the experimental configuration described in fig. \ref{fig:fig1}c, the collection polarization is set orthogonal to the excitation polarization by the PBS, hence in the $H_\text{coll}$ polarization direction (see fig. ~\ref{fig:fig3}a). 
For $\phi=0$, the excitation is parallel to the $V$ cavity axis and only the spectrally narrow emission arising from the QD is collected in the $H$-mode, which corresponds to the configuration used for the source operation. However, when $\phi$ is different from $0\deg$ or $90\deg$, an additional broader emission line is visible as shown in fig. ~\ref{fig:fig3}b for the two devices S5 and S13. This broader emission arises from the  birefringence of the cavity, as explained in subsection \ref{neutral_exciton_sources}. For better visualization of the cavity-induced light rotation, we used 3 ps laser pulses, which are spectrally broader than the cavity linewidth. The emission of the cavity rotated light is maximal for around $\phi=45^{\circ} $ and $ 135^{\circ}$ \cite{hilaire_accurate_2018}.

Fig. ~\ref{fig:fig3}c shows the emission spectra measured for the exciton S5,  as a function of $\phi$.  Fig. ~\ref{fig:fig3}e presents the intensity of the cavity rotated light and the QD emission as a function of $\phi$. The emission arising from the exciton has a sinusoidal dependance with $\phi$. A laser polarized along one of the exciton axes $V'$ or $H'$ ($\theta=0$ or $\pi/2$) excites an eigenstate of the system and no  emission takes place in the orthogonal polarization. The emission of the exciton thus depends on the angle between the incident polarization and the exciton axes ($\theta-\phi$). This measurement also allows to determine $\theta$, the angle between the cavity polarization and the exciton axes, an important parameter for the source brightness (see eq. (\ref{eq_Hproj})).
Fig. ~\ref{fig:fig3}d and 3f show the same experiment and analysis on a trion-based source S13. The rotated light arising from the cavity has the same sinusoidal dependence but the trion emission is now independent of $\phi$. Indeed, the selection rules illustrated in fig. ~\ref{fig:fig2}c can be rewritten in any two-orthogonal linear polarization basis and the amplitude of the QD emission in the polarization orthogonal to the excitation is independent of the orientation of the excitation.

\section{Source benchmarking }
\label{benchmarking}

We now discuss the benchmarking of fifteen devices, whose main quantum properties are presented in fig. ~\ref{fig:fig4}. We consider not only the figures of merit defining each source performance (single-photon purity, indistinguishability and brightness), but also the emission wavelength and temporal profile, that are critical characteristics for large-scale fabrication of identical sources. All figures of merit are reported using $\pi$-pulse excitation~\cite{Stievater2001RabiOO} with $15$~ps pulses at a repetition rate of 81 MHz. 

Fig. ~\ref{fig:fig4}a shows the  $g^{(2)}(0)$ values of the sources, with an average of $4.6 \pm 1.5\% $. The single-photon purity ($1{-}g^{(2)}(0)$) is systematically  higher for excitons than for trion sources. This can be understood by comparing the different emission processes for each source type in cross-polarization (shown in fig. ~\ref{fig:fig2}b and fig. ~\ref{fig:fig2}d). Since the emission from the exciton through the $H$ mode of the cavity is delayed compared to the excitation pulse, the probability that the quantum dot gets re-excited afterwards is low because the excitation pulse is over. In the case of a trion, the emission process in cross-polarization begins on the same timescale as the pulse and so there is a higher probability of re-excitation of the transition within the same excitation pulse, leading to the emission of a second photon~\cite{fischer_signatures_2017,loredo_generation_2019}.

Fig. ~\ref{fig:fig4}b shows the mean wavepacket overlap between two consecutive single photons emitted with a 12 ns delay. These values are obtained from the HOM visibility histogram and corrected from the corresponding non-zero $g^{(2)}(0)$~\cite{g2HOM}. The mean wavepacket overlap over the various sources is quite homogeneous, with an average value of $90.6 \pm 2.8\% $. Note that this indistinguishability is obtained without spectral filtering and thus it includes the contribution from the phonon sideband, which is strongly suppressed by the cavity funneling effect ~\cite{grange_reducing_2017,iles-smith_phonon_2017}. The slightly lower indistinguishability values of the sources from sample E are likely due to a higher temperature on that chip (around 10-11 K). These observations also show that, despite a very different temporal structure of the single-photon wavepacket in the case of exciton based sources, the coherence is highly preserved in the frequency domain over the emission process.

The source brightness corresponds to the probability to obtain a single photon for a given excitation pulse. It is defined as the collected single-photon rate divided by the excitation repetition rate~\cite{senellart_high-performance_2017}.  This probability is measured at the output of the collection fiber and then deduced at the first lens (fig. ~\ref{fig:fig4}c). Both values depend on the global efficiency of our setup (transmission and output coupling), which is approximately $40\%$. The average first lens brightness is $13.6 \pm  4.4 \%$, which is on par with state-of-the-art values using a 0.45 NA collection objective.

In a cross-polarization setup using cavities with a small birefringence, both the first lens and fibered brightness are limited to at most $50\%$ due to the rejection of photons orthogonally polarized to the collection mode. However, this limit can be overcome using polarized cavities \cite{wang_towards_2019}. In addition, the brightness is limited by the extraction efficiency, which is intrinsic to the cavity and is very similar for all pillars ($\sim 65\%$), and also by the occupation probability of the QD charge state. Indeed, the rather large variation in brightness among the trion sources could be reduced by a  better control on the occupation probability of the hole in the QD~\cite{hilaire_2019}. The brightness of an exciton based source further depends on both $\theta$ and the relative values of $\Delta_\text{FSS}$ and the emission rate $1/\tau$. Because the values of $\theta$ and FSS are not controlled during fabrication, excitons tend to be slightly less bright than trions. 

The table in fig. ~\ref{fig:fig4}e presents the fibered single-photon rate obtained by dividing the detected count rates measured at the output of a single-mode fiber by the efficiency  of our silicon based avalanche photodiode ($30\%$ detection efficiency at ${\sim}925$ nm). 
\noindent It is also worth noting that we use a long-distance microscope objective with a numerical aperture of 0.45 that is placed outside of the cryostat. This limits the brightness because not all of the emission is collected by the objective. Also, this spatial truncation of the beam leads to a deformation of the wavefront which reduces the coupling into the single-mode fiber of typically 50 to 60\%. Such collection efficiency could be strongly improved by inserting a high numerical aperture lens into the cryostat itself.

 \vspace{0.5cm}
 
\section{Perspectives for scalability}
\label{perspectives}

Our systematic study of a large number of single-photon sources allows identification of the remaining challenges for the fabrication of remote sources generating highly identical photons. 
\noindent In this context, a  key parameter is the source operation wavelength. The typical  inhomogeneous broadening of the InGaAs QDs spectum is around $30$ nm. However, the in-situ lithography allows us to select QDs in a small spectral range and to fabricate pillars with the correct diameter to match the QD resonance~\cite{nowak_deterministic_2014}. Fig. ~\ref{fig:fig4}d shows the optimal operation wavelength of the sources. Overall, a  small deviation of the operation wavelength is observed with an average wavelength of 924.7~nm with a standard deviation of only $0.5\ $~nm.  When considering sources fabricated on the same part of the original wafer, this deviation is substantially reduced with $0.06\ $~nm for D and $0.12\ $~nm for E, showing the high degree of control provided by the in-situ lithography. Thus, we conclude that a higher homogeneity in the growth process across the wafer, as typically obtained from industrial epitaxy growth systems, would increase the yield of sources operating at the same wavelength. Note that the residual $0.06\ $~nm of deviation can be compensated via Stark tuning with minimal reduction to the brightness considering that the typical linewidth of our micropillar cavities is $0.2\ $~nm. 

The temporal profile of the single-photon wavepackets is also another important feature to consider. We show that the challenges are different for sources based on neutral or charged  QDs in a cross-polarized setup. Controlling the temporal profile of a exciton based source is challenging because it requires control of the fine-structure splitting and the relative axes orientation of the cavity and QD. Such complete control was not achieved here, but many tools have been developed in the last few years to control both the exciton fine-structure splitting~\cite{trotta_energy-tunable_2015} and the cavity birefringence~\cite{wang_towards_2019}. On the other hand,  we find that the temporal profiles of the trion-based sources are very consistent across all samples, with an average decay time of $180$~ps and a standard deviation of only $17$~ps (see fig. ~\ref{fig:fig4}f).   

\noindent The most promising path to scalability thus lies in the development of trion based sources where higher single-photon purity can be obtained by using shorter excitation pulses to reduce the probability for re-excitation. Such short pulses have not been used here due to the chromatic dependence of the polarization rejection apparatus. Using a specially designed highly achromatic polarization control or a side excitation scheme~\cite{weihs_deterministic_2012} would improve of the sources performance. Finally, we note that the variation in the brightness of the trion-based sources is rather large. This is due to an imperfect control of the QD charge state. This could be improved by applying an electric field during the in-situ lithography process, which would allow us to explore the different available charge states and their characteristics prior to the QD selection and cavity definition.

\section{Conclusion}
\label{conclusion}
We have benchmarked a large number of single-photon sources based on semiconductor QDs.  The present technology is based on the deliberate choice of using naturally grown QDs, which currently have a higher quantum purity than site controlled QDs~\cite{jons_triggered_2013,PhysRevB.93.195316}.  Despite the QD random distribution, in both spatial and spectral degrees of freedom, our technology based on selecting the QDs during an in-situ-lithography step and building a tailored cavity around them, allows us to obtain a large number of sources with highly homogeneous properties. The first-lens brightness is shown to reach state-of-the-art values for resonant excitation with a highly reproducible indistinguishability of $90.6 \pm 2.8\% $. Moreover, we discussed the physics of the sources behaviour in a cross-polarized resonant excitation scheme, and identified the parameters controlling the wavepacket temporal profile,  the source brightness and the single-photon purity. The present study shows a clear path for scaling the fabrication process where highly identical remote sources could be obtained using a trion. 

 \vspace{0.5cm}

\begin{acknowledgments}
This work was partially supported by the ERC PoC PhoW, the French Agence Nationale pour la Recherche (grant ANR SPIQE, USSEPP, QuDICE), the IAD - ANR support ASTRID program Projet ANR-18-ASTR-0024 LIGHT,  the QuantERA ERA-NET Cofund in Quantum Technologies, project HIPHOP, the French RENATECH network, a public grant overseen by the French National Research Agency (ANR) as part of the "Investissements d'Avenir" programme (Labex NanoSaclay, reference: ANR-10-LABX-0035). J.C.L. and C.A. acknowledge support from Marie Sk\l{}odowska-Curie Individual Fellowships SMUPHOS and SQUAPH, respectively. H. O. Acknowledges support from  the Paris Ile-de-France Région in the framework of DIM SIRTEQ.

\end{acknowledgments}

\bibliographystyle{apsrev4-1}
 
\bibliography{HOllivier2019}

\begin{thebibliography}{52}%
\makeatletter
\providecommand \@ifxundefined [1]{%
 \@ifx{#1\undefined}
}%
\providecommand \@ifnum [1]{%
 \ifnum #1\expandafter \@firstoftwo
 \else \expandafter \@secondoftwo
 \fi
}%
\providecommand \@ifx [1]{%
 \ifx #1\expandafter \@firstoftwo
 \else \expandafter \@secondoftwo
 \fi
}%
\providecommand \natexlab [1]{#1}%
\providecommand \enquote  [1]{``#1''}%
\providecommand \bibnamefont  [1]{#1}%
\providecommand \bibfnamefont [1]{#1}%
\providecommand \citenamefont [1]{#1}%
\providecommand \href@noop [0]{\@secondoftwo}%
\providecommand \href [0]{\begingroup \@sanitize@url \@href}%
\providecommand \@href[1]{\@@startlink{#1}\@@href}%
\providecommand \@@href[1]{\endgroup#1\@@endlink}%
\providecommand \@sanitize@url [0]{\catcode `\\12\catcode `\$12\catcode
  `\&12\catcode `\#12\catcode `\^12\catcode `\_12\catcode `\%12\relax}%
\providecommand \@@startlink[1]{}%
\providecommand \@@endlink[0]{}%
\providecommand \url  [0]{\begingroup\@sanitize@url \@url }%
\providecommand \@url [1]{\endgroup\@href {#1}{\urlprefix }}%
\providecommand \urlprefix  [0]{URL }%
\providecommand \Eprint [0]{\href }%
\providecommand \doibase [0]{http://dx.doi.org/}%
\providecommand \selectlanguage [0]{\@gobble}%
\providecommand \bibinfo  [0]{\@secondoftwo}%
\providecommand \bibfield  [0]{\@secondoftwo}%
\providecommand \translation [1]{[#1]}%
\providecommand \BibitemOpen [0]{}%
\providecommand \bibitemStop [0]{}%
\providecommand \bibitemNoStop [0]{.\EOS\space}%
\providecommand \EOS [0]{\spacefactor3000\relax}%
\providecommand \BibitemShut  [1]{\csname bibitem#1\endcsname}%
\let\auto@bib@innerbib\@empty
\bibitem [{\citenamefont {M{\'a}ttar}\ \emph {et~al.}(2018)\citenamefont
  {M{\'a}ttar}, \citenamefont {Ko{\l}ody{\'n}ski}, \citenamefont {Skrzypczyk},
  \citenamefont {Cavalcanti}, \citenamefont {Banaszek},\ and\ \citenamefont
  {Ac{\'\i}n}}]{mattar_device-independent_2018}%
  \BibitemOpen
  \bibfield  {author} {\bibinfo {author} {\bibfnamefont {A.}~\bibnamefont
  {M{\'a}ttar}}, \bibinfo {author} {\bibfnamefont {J.}~\bibnamefont
  {Ko{\l}ody{\'n}ski}}, \bibinfo {author} {\bibfnamefont {P.}~\bibnamefont
  {Skrzypczyk}}, \bibinfo {author} {\bibfnamefont {D.}~\bibnamefont
  {Cavalcanti}}, \bibinfo {author} {\bibfnamefont {K.}~\bibnamefont
  {Banaszek}}, \ and\ \bibinfo {author} {\bibfnamefont {A.}~\bibnamefont
  {Ac{\'\i}n}},\ }\href {http://arxiv.org/abs/1803.07089} {\bibfield  {journal}
  {\bibinfo  {journal} {arXiv:1803.07089 [quant-ph]}\ } (\bibinfo {year}
  {2018})},\ \bibinfo {note} {arXiv: 1803.07089}\BibitemShut {NoStop}%
\bibitem [{\citenamefont {Hensen}\ \emph {et~al.}(2015)\citenamefont {Hensen},
  \citenamefont {Bernien}, \citenamefont {Dr{\'e}au}, \citenamefont {Reiserer},
  \citenamefont {Kalb}, \citenamefont {Blok}, \citenamefont {Ruitenberg},
  \citenamefont {Vermeulen}, \citenamefont {Schouten}, \citenamefont
  {Abell{\'a}n}, \citenamefont {Amaya}, \citenamefont {Pruneri}, \citenamefont
  {Mitchell}, \citenamefont {Markham}, \citenamefont {Twitchen}, \citenamefont
  {Elkouss}, \citenamefont {Wehner}, \citenamefont {Taminiau},\ and\
  \citenamefont {Hanson}}]{hensen_loophole-free_2015}%
  \BibitemOpen
  \bibfield  {author} {\bibinfo {author} {\bibfnamefont {B.}~\bibnamefont
  {Hensen}}, \bibinfo {author} {\bibfnamefont {H.}~\bibnamefont {Bernien}},
  \bibinfo {author} {\bibfnamefont {A.~E.}\ \bibnamefont {Dr{\'e}au}}, \bibinfo
  {author} {\bibfnamefont {A.}~\bibnamefont {Reiserer}}, \bibinfo {author}
  {\bibfnamefont {N.}~\bibnamefont {Kalb}}, \bibinfo {author} {\bibfnamefont
  {M.~S.}\ \bibnamefont {Blok}}, \bibinfo {author} {\bibfnamefont
  {J.}~\bibnamefont {Ruitenberg}}, \bibinfo {author} {\bibfnamefont {R.~F.~L.}\
  \bibnamefont {Vermeulen}}, \bibinfo {author} {\bibfnamefont {R.~N.}\
  \bibnamefont {Schouten}}, \bibinfo {author} {\bibfnamefont {C.}~\bibnamefont
  {Abell{\'a}n}}, \bibinfo {author} {\bibfnamefont {W.}~\bibnamefont {Amaya}},
  \bibinfo {author} {\bibfnamefont {V.}~\bibnamefont {Pruneri}}, \bibinfo
  {author} {\bibfnamefont {M.~W.}\ \bibnamefont {Mitchell}}, \bibinfo {author}
  {\bibfnamefont {M.}~\bibnamefont {Markham}}, \bibinfo {author} {\bibfnamefont
  {D.~J.}\ \bibnamefont {Twitchen}}, \bibinfo {author} {\bibfnamefont
  {D.}~\bibnamefont {Elkouss}}, \bibinfo {author} {\bibfnamefont
  {S.}~\bibnamefont {Wehner}}, \bibinfo {author} {\bibfnamefont {T.~H.}\
  \bibnamefont {Taminiau}}, \ and\ \bibinfo {author} {\bibfnamefont
  {R.}~\bibnamefont {Hanson}},\ }\href {\doibase 10.1038/nature15759}
  {\bibfield  {journal} {\bibinfo  {journal} {Nature}\ }\textbf {\bibinfo
  {volume} {526}},\ \bibinfo {pages} {682} (\bibinfo {year}
  {2015})}\BibitemShut {NoStop}%
\bibitem [{\citenamefont {Liao}\ \emph {et~al.}(2017)\citenamefont {Liao},
  \citenamefont {qi~Cai}, \citenamefont {Liu}, \citenamefont {Zhang},
  \citenamefont {Li}, \citenamefont {Ren}, \citenamefont {Yin}, \citenamefont
  {Shen}, \citenamefont {Cao}, \citenamefont {Li}, \citenamefont {Li},
  \citenamefont {Chen}, \citenamefont {hua Sun}, \citenamefont {Jia},
  \citenamefont {Wu}, \citenamefont {jun Jiang}, \citenamefont {feng Wang},
  \citenamefont {mei Huang}, \citenamefont {Wang}, \citenamefont {Zhou},
  \citenamefont {Deng}, \citenamefont {Xi}, \citenamefont {Ma}, \citenamefont
  {Hu}, \citenamefont {Zhang}, \citenamefont {Chen}, \citenamefont {Liu},
  \citenamefont {Wang}, \citenamefont {cai Zhu}, \citenamefont {Lu},
  \citenamefont {Shu}, \citenamefont {Peng}, \citenamefont {Wang},\ and\
  \citenamefont {Pan}}]{Liao2017SatellitetogroundQK}%
  \BibitemOpen
  \bibfield  {author} {\bibinfo {author} {\bibfnamefont {S.-K.}\ \bibnamefont
  {Liao}}, \bibinfo {author} {\bibfnamefont {W.}~\bibnamefont {qi~Cai}},
  \bibinfo {author} {\bibfnamefont {W.-Y.}\ \bibnamefont {Liu}}, \bibinfo
  {author} {\bibfnamefont {J.}~\bibnamefont {Zhang}}, \bibinfo {author}
  {\bibfnamefont {Y.}~\bibnamefont {Li}}, \bibinfo {author} {\bibfnamefont
  {J.-G.}\ \bibnamefont {Ren}}, \bibinfo {author} {\bibfnamefont
  {J.}~\bibnamefont {Yin}}, \bibinfo {author} {\bibfnamefont {Q.}~\bibnamefont
  {Shen}}, \bibinfo {author} {\bibfnamefont {Y.}~\bibnamefont {Cao}}, \bibinfo
  {author} {\bibfnamefont {Z.-P.}\ \bibnamefont {Li}}, \bibinfo {author}
  {\bibfnamefont {F.-Z.}\ \bibnamefont {Li}}, \bibinfo {author} {\bibfnamefont
  {X.-W.}\ \bibnamefont {Chen}}, \bibinfo {author} {\bibfnamefont
  {L.}~\bibnamefont {hua Sun}}, \bibinfo {author} {\bibfnamefont {J.-J.}\
  \bibnamefont {Jia}}, \bibinfo {author} {\bibfnamefont {J.-C.}\ \bibnamefont
  {Wu}}, \bibinfo {author} {\bibfnamefont {X.}~\bibnamefont {jun Jiang}},
  \bibinfo {author} {\bibfnamefont {J.}~\bibnamefont {feng Wang}}, \bibinfo
  {author} {\bibfnamefont {Y.}~\bibnamefont {mei Huang}}, \bibinfo {author}
  {\bibfnamefont {Q.}~\bibnamefont {Wang}}, \bibinfo {author} {\bibfnamefont
  {Y.-L.}\ \bibnamefont {Zhou}}, \bibinfo {author} {\bibfnamefont
  {L.}~\bibnamefont {Deng}}, \bibinfo {author} {\bibfnamefont {T.}~\bibnamefont
  {Xi}}, \bibinfo {author} {\bibfnamefont {L.}~\bibnamefont {Ma}}, \bibinfo
  {author} {\bibfnamefont {T.}~\bibnamefont {Hu}}, \bibinfo {author}
  {\bibfnamefont {Q.}~\bibnamefont {Zhang}}, \bibinfo {author} {\bibfnamefont
  {Y.-A.}\ \bibnamefont {Chen}}, \bibinfo {author} {\bibfnamefont {N.-L.}\
  \bibnamefont {Liu}}, \bibinfo {author} {\bibfnamefont {X.-B.}\ \bibnamefont
  {Wang}}, \bibinfo {author} {\bibfnamefont {Z.}~\bibnamefont {cai Zhu}},
  \bibinfo {author} {\bibfnamefont {C.-Y.}\ \bibnamefont {Lu}}, \bibinfo
  {author} {\bibfnamefont {R.}~\bibnamefont {Shu}}, \bibinfo {author}
  {\bibfnamefont {C.-Z.}\ \bibnamefont {Peng}}, \bibinfo {author}
  {\bibfnamefont {J.-Y.}\ \bibnamefont {Wang}}, \ and\ \bibinfo {author}
  {\bibfnamefont {J.-W.}\ \bibnamefont {Pan}},\ }\href@noop {} {\bibfield
  {journal} {\bibinfo  {journal} {Nature}\ }\textbf {\bibinfo {volume} {549}},\
  \bibinfo {pages} {43} (\bibinfo {year} {2017})}\BibitemShut {NoStop}%
\bibitem [{\citenamefont {Taballione}\ \emph {et~al.}(2019)\citenamefont
  {Taballione}, \citenamefont {Wolterink}, \citenamefont {Lugani},
  \citenamefont {Eckstein}, \citenamefont {Bell}, \citenamefont {Grootjans},
  \citenamefont {Visscher}, \citenamefont {Geskus}, \citenamefont {Roeloffzen},
  \citenamefont {Renema}, \citenamefont {Walmsley}, \citenamefont {Pinkse},\
  and\ \citenamefont {Boller}}]{taballione_88_2019}%
  \BibitemOpen
  \bibfield  {author} {\bibinfo {author} {\bibfnamefont {C.}~\bibnamefont
  {Taballione}}, \bibinfo {author} {\bibfnamefont {T.~A.~W.}\ \bibnamefont
  {Wolterink}}, \bibinfo {author} {\bibfnamefont {J.}~\bibnamefont {Lugani}},
  \bibinfo {author} {\bibfnamefont {A.}~\bibnamefont {Eckstein}}, \bibinfo
  {author} {\bibfnamefont {B.~A.}\ \bibnamefont {Bell}}, \bibinfo {author}
  {\bibfnamefont {R.}~\bibnamefont {Grootjans}}, \bibinfo {author}
  {\bibfnamefont {I.}~\bibnamefont {Visscher}}, \bibinfo {author}
  {\bibfnamefont {D.}~\bibnamefont {Geskus}}, \bibinfo {author} {\bibfnamefont
  {C.~G.~H.}\ \bibnamefont {Roeloffzen}}, \bibinfo {author} {\bibfnamefont
  {J.~J.}\ \bibnamefont {Renema}}, \bibinfo {author} {\bibfnamefont {I.~A.}\
  \bibnamefont {Walmsley}}, \bibinfo {author} {\bibfnamefont {P.~W.~H.}\
  \bibnamefont {Pinkse}}, \ and\ \bibinfo {author} {\bibfnamefont {K.-J.}\
  \bibnamefont {Boller}},\ }\href {\doibase 10.1364/OE.27.026842} {\bibfield
  {journal} {\bibinfo  {journal} {Opt. Express}\ }\textbf {\bibinfo {volume}
  {27}},\ \bibinfo {pages} {26842} (\bibinfo {year} {2019})}\BibitemShut
  {NoStop}%
\bibitem [{\citenamefont {Wang}\ \emph {et~al.}(2018)\citenamefont {Wang},
  \citenamefont {Paesani}, \citenamefont {Ding}, \citenamefont {Santagati},
  \citenamefont {Skrzypczyk}, \citenamefont {Salavrakos}, \citenamefont
  {Tura~Brugu{\'e}s}, \citenamefont {Augusiak}, \citenamefont {Mancinska},
  \citenamefont {Bacco}, \citenamefont {Bonneau}, \citenamefont {Silverstone},
  \citenamefont {Gong}, \citenamefont {Ac{\'\i}n}, \citenamefont {Rottwitt},
  \citenamefont {Oxenl{\o}we}, \citenamefont {O'Brien}, \citenamefont {Laing},\
  and\ \citenamefont {Thompson}}]{wang_multidimensional_2018}%
  \BibitemOpen
  \bibfield  {author} {\bibinfo {author} {\bibfnamefont {J.}~\bibnamefont
  {Wang}}, \bibinfo {author} {\bibfnamefont {S.}~\bibnamefont {Paesani}},
  \bibinfo {author} {\bibfnamefont {Y.}~\bibnamefont {Ding}}, \bibinfo {author}
  {\bibfnamefont {R.}~\bibnamefont {Santagati}}, \bibinfo {author}
  {\bibfnamefont {P.}~\bibnamefont {Skrzypczyk}}, \bibinfo {author}
  {\bibfnamefont {A.}~\bibnamefont {Salavrakos}}, \bibinfo {author}
  {\bibfnamefont {J.}~\bibnamefont {Tura~Brugu{\'e}s}}, \bibinfo {author}
  {\bibfnamefont {R.}~\bibnamefont {Augusiak}}, \bibinfo {author}
  {\bibfnamefont {L.}~\bibnamefont {Mancinska}}, \bibinfo {author}
  {\bibfnamefont {D.}~\bibnamefont {Bacco}}, \bibinfo {author} {\bibfnamefont
  {D.}~\bibnamefont {Bonneau}}, \bibinfo {author} {\bibfnamefont
  {J.}~\bibnamefont {Silverstone}}, \bibinfo {author} {\bibfnamefont
  {Q.}~\bibnamefont {Gong}}, \bibinfo {author} {\bibfnamefont {A.}~\bibnamefont
  {Ac{\'\i}n}}, \bibinfo {author} {\bibfnamefont {K.}~\bibnamefont {Rottwitt}},
  \bibinfo {author} {\bibfnamefont {L.}~\bibnamefont {Oxenl{\o}we}}, \bibinfo
  {author} {\bibfnamefont {J.}~\bibnamefont {O'Brien}}, \bibinfo {author}
  {\bibfnamefont {A.}~\bibnamefont {Laing}}, \ and\ \bibinfo {author}
  {\bibfnamefont {M.}~\bibnamefont {Thompson}},\ }\href {\doibase
  10.1126/science.aar7053} {\bibfield  {journal} {\bibinfo  {journal} {Science
  (New York, N.Y.)}\ }\textbf {\bibinfo {volume} {360}} (\bibinfo {year}
  {2018}),\ 10.1126/science.aar7053}\BibitemShut {NoStop}%
\bibitem [{\citenamefont {Hosseini}\ \emph {et~al.}(2011)\citenamefont
  {Hosseini}, \citenamefont {Campbell}, \citenamefont {Sparkes}, \citenamefont
  {Lam},\ and\ \citenamefont {Buchler}}]{hosseini_unconditional_2011}%
  \BibitemOpen
  \bibfield  {author} {\bibinfo {author} {\bibfnamefont {M.}~\bibnamefont
  {Hosseini}}, \bibinfo {author} {\bibfnamefont {G.}~\bibnamefont {Campbell}},
  \bibinfo {author} {\bibfnamefont {B.~M.}\ \bibnamefont {Sparkes}}, \bibinfo
  {author} {\bibfnamefont {P.~K.}\ \bibnamefont {Lam}}, \ and\ \bibinfo
  {author} {\bibfnamefont {B.~C.}\ \bibnamefont {Buchler}},\ }\href
  {https://doi.org/10.1038/nphys2021} {\bibfield  {journal} {\bibinfo
  {journal} {Nature Physics}\ }\textbf {\bibinfo {volume} {7}},\ \bibinfo
  {pages} {794} (\bibinfo {year} {2011})}\BibitemShut {NoStop}%
\bibitem [{\citenamefont {Brida}\ \emph {et~al.}(2010)\citenamefont {Brida},
  \citenamefont {Genovese},\ and\ \citenamefont
  {Ruo~Berchera}}]{brida_experimental_2010}%
  \BibitemOpen
  \bibfield  {author} {\bibinfo {author} {\bibfnamefont {G.}~\bibnamefont
  {Brida}}, \bibinfo {author} {\bibfnamefont {M.}~\bibnamefont {Genovese}}, \
  and\ \bibinfo {author} {\bibfnamefont {I.}~\bibnamefont {Ruo~Berchera}},\
  }\href {https://doi.org/10.1038/nphoton.2010.29} {\bibfield  {journal}
  {\bibinfo  {journal} {Nature Photonics}\ }\textbf {\bibinfo {volume} {4}},\
  \bibinfo {pages} {227} (\bibinfo {year} {2010})}\BibitemShut {NoStop}%
\bibitem [{\citenamefont {Forbes}\ and\ \citenamefont
  {Rodriguez-Fajardo}(2019)}]{forbes_super-resolution_2019}%
  \BibitemOpen
  \bibfield  {author} {\bibinfo {author} {\bibfnamefont {A.}~\bibnamefont
  {Forbes}}\ and\ \bibinfo {author} {\bibfnamefont {V.}~\bibnamefont
  {Rodriguez-Fajardo}},\ }\href {\doibase 10.1038/s41566-018-0344-8} {\bibfield
   {journal} {\bibinfo  {journal} {Nature Photonics}\ }\textbf {\bibinfo
  {volume} {13}},\ \bibinfo {pages} {76} (\bibinfo {year} {2019})}\BibitemShut
  {NoStop}%
\bibitem [{\citenamefont {Mosley}\ \emph {et~al.}(2008)\citenamefont {Mosley},
  \citenamefont {Lundeen}, \citenamefont {Smith}, \citenamefont {Wasylczyk},
  \citenamefont {U'Ren}, \citenamefont {Silberhorn},\ and\ \citenamefont
  {Walmsley}}]{mosley_heralded_2008}%
  \BibitemOpen
  \bibfield  {author} {\bibinfo {author} {\bibfnamefont {P.~J.}\ \bibnamefont
  {Mosley}}, \bibinfo {author} {\bibfnamefont {J.~S.}\ \bibnamefont {Lundeen}},
  \bibinfo {author} {\bibfnamefont {B.~J.}\ \bibnamefont {Smith}}, \bibinfo
  {author} {\bibfnamefont {P.}~\bibnamefont {Wasylczyk}}, \bibinfo {author}
  {\bibfnamefont {A.~B.}\ \bibnamefont {U'Ren}}, \bibinfo {author}
  {\bibfnamefont {C.}~\bibnamefont {Silberhorn}}, \ and\ \bibinfo {author}
  {\bibfnamefont {I.~A.}\ \bibnamefont {Walmsley}},\ }\href {\doibase
  10.1103/PhysRevLett.100.133601} {\bibfield  {journal} {\bibinfo  {journal}
  {Phys. Rev. Lett.}\ }\textbf {\bibinfo {volume} {100}},\ \bibinfo {pages}
  {133601} (\bibinfo {year} {2008})}\BibitemShut {NoStop}%
\bibitem [{\citenamefont {Kaneda}\ \emph {et~al.}(2015)\citenamefont {Kaneda},
  \citenamefont {Christensen}, \citenamefont {Wong}, \citenamefont {Park},
  \citenamefont {McCusker},\ and\ \citenamefont
  {Kwiat}}]{kaneda_time-multiplexed_2015}%
  \BibitemOpen
  \bibfield  {author} {\bibinfo {author} {\bibfnamefont {F.}~\bibnamefont
  {Kaneda}}, \bibinfo {author} {\bibfnamefont {B.~G.}\ \bibnamefont
  {Christensen}}, \bibinfo {author} {\bibfnamefont {J.~J.}\ \bibnamefont
  {Wong}}, \bibinfo {author} {\bibfnamefont {H.~S.}\ \bibnamefont {Park}},
  \bibinfo {author} {\bibfnamefont {K.~T.}\ \bibnamefont {McCusker}}, \ and\
  \bibinfo {author} {\bibfnamefont {P.~G.}\ \bibnamefont {Kwiat}},\ }\href
  {\doibase 10.1364/OPTICA.2.001010} {\bibfield  {journal} {\bibinfo  {journal}
  {Optica}\ }\textbf {\bibinfo {volume} {2}},\ \bibinfo {pages} {1010}
  (\bibinfo {year} {2015})}\BibitemShut {NoStop}%
\bibitem [{\citenamefont {Kaneda}\ and\ \citenamefont
  {Kwiat}(2018)}]{kaneda_high-efficiency_2018}%
  \BibitemOpen
  \bibfield  {author} {\bibinfo {author} {\bibfnamefont {F.}~\bibnamefont
  {Kaneda}}\ and\ \bibinfo {author} {\bibfnamefont {P.~G.}\ \bibnamefont
  {Kwiat}},\ }\href {http://arxiv.org/abs/1803.04803} {\bibfield  {journal}
  {\bibinfo  {journal} {arXiv:1803.04803 [quant-ph]}\ } (\bibinfo {year}
  {2018})},\ \bibinfo {note} {arXiv: 1803.04803}\BibitemShut {NoStop}%
\bibitem [{\citenamefont {Somaschi}\ \emph {et~al.}(2016)\citenamefont
  {Somaschi}, \citenamefont {Giesz}, \citenamefont {De~Santis}, \citenamefont
  {Loredo}, \citenamefont {Almeida}, \citenamefont {Hornecker}, \citenamefont
  {Portalupi}, \citenamefont {Grange}, \citenamefont {Ant{\'o}n}, \citenamefont
  {Demory}, \citenamefont {G{\'o}mez}, \citenamefont {Sagnes}, \citenamefont
  {Lanzillotti-Kimura}, \citenamefont {Lema{\'\i}tre}, \citenamefont
  {Auffeves}, \citenamefont {White}, \citenamefont {Lanco},\ and\ \citenamefont
  {Senellart}}]{somaschi_near-optimal_2016}%
  \BibitemOpen
  \bibfield  {author} {\bibinfo {author} {\bibfnamefont {N.}~\bibnamefont
  {Somaschi}}, \bibinfo {author} {\bibfnamefont {V.}~\bibnamefont {Giesz}},
  \bibinfo {author} {\bibfnamefont {L.}~\bibnamefont {De~Santis}}, \bibinfo
  {author} {\bibfnamefont {J.~C.}\ \bibnamefont {Loredo}}, \bibinfo {author}
  {\bibfnamefont {M.~P.}\ \bibnamefont {Almeida}}, \bibinfo {author}
  {\bibfnamefont {G.}~\bibnamefont {Hornecker}}, \bibinfo {author}
  {\bibfnamefont {S.~L.}\ \bibnamefont {Portalupi}}, \bibinfo {author}
  {\bibfnamefont {T.}~\bibnamefont {Grange}}, \bibinfo {author} {\bibfnamefont
  {C.}~\bibnamefont {Ant{\'o}n}}, \bibinfo {author} {\bibfnamefont
  {J.}~\bibnamefont {Demory}}, \bibinfo {author} {\bibfnamefont
  {C.}~\bibnamefont {G{\'o}mez}}, \bibinfo {author} {\bibfnamefont
  {I.}~\bibnamefont {Sagnes}}, \bibinfo {author} {\bibfnamefont {N.~D.}\
  \bibnamefont {Lanzillotti-Kimura}}, \bibinfo {author} {\bibfnamefont
  {A.}~\bibnamefont {Lema{\'\i}tre}}, \bibinfo {author} {\bibfnamefont
  {A.}~\bibnamefont {Auffeves}}, \bibinfo {author} {\bibfnamefont {A.~G.}\
  \bibnamefont {White}}, \bibinfo {author} {\bibfnamefont {L.}~\bibnamefont
  {Lanco}}, \ and\ \bibinfo {author} {\bibfnamefont {P.}~\bibnamefont
  {Senellart}},\ }\href {https://doi.org/10.1038/nphoton.2016.23} {\bibfield
  {journal} {\bibinfo  {journal} {Nature Photonics}\ }\textbf {\bibinfo
  {volume} {10}},\ \bibinfo {pages} {340} (\bibinfo {year} {2016})}\BibitemShut
  {NoStop}%
\bibitem [{\citenamefont {Wang}\ \emph {et~al.}(2019)\citenamefont {Wang},
  \citenamefont {He}, \citenamefont {Chung}, \citenamefont {Hu}, \citenamefont
  {Yu}, \citenamefont {Chen}, \citenamefont {Ding}, \citenamefont {Chen},
  \citenamefont {Qin}, \citenamefont {Yang}, \citenamefont {Liu}, \citenamefont
  {Duan}, \citenamefont {Li}, \citenamefont {Gerhardt}, \citenamefont
  {Winkler}, \citenamefont {Jurkat}, \citenamefont {Wang}, \citenamefont
  {Gregersen}, \citenamefont {Huo}, \citenamefont {Dai}, \citenamefont {Yu},
  \citenamefont {H{\"o}fling}, \citenamefont {Lu},\ and\ \citenamefont
  {Pan}}]{wang_towards_2019}%
  \BibitemOpen
  \bibfield  {author} {\bibinfo {author} {\bibfnamefont {H.}~\bibnamefont
  {Wang}}, \bibinfo {author} {\bibfnamefont {Y.-M.}\ \bibnamefont {He}},
  \bibinfo {author} {\bibfnamefont {T.-H.}\ \bibnamefont {Chung}}, \bibinfo
  {author} {\bibfnamefont {H.}~\bibnamefont {Hu}}, \bibinfo {author}
  {\bibfnamefont {Y.}~\bibnamefont {Yu}}, \bibinfo {author} {\bibfnamefont
  {S.}~\bibnamefont {Chen}}, \bibinfo {author} {\bibfnamefont {X.}~\bibnamefont
  {Ding}}, \bibinfo {author} {\bibfnamefont {M.-C.}\ \bibnamefont {Chen}},
  \bibinfo {author} {\bibfnamefont {J.}~\bibnamefont {Qin}}, \bibinfo {author}
  {\bibfnamefont {X.}~\bibnamefont {Yang}}, \bibinfo {author} {\bibfnamefont
  {R.-Z.}\ \bibnamefont {Liu}}, \bibinfo {author} {\bibfnamefont {Z.-C.}\
  \bibnamefont {Duan}}, \bibinfo {author} {\bibfnamefont {J.-P.}\ \bibnamefont
  {Li}}, \bibinfo {author} {\bibfnamefont {S.}~\bibnamefont {Gerhardt}},
  \bibinfo {author} {\bibfnamefont {K.}~\bibnamefont {Winkler}}, \bibinfo
  {author} {\bibfnamefont {J.}~\bibnamefont {Jurkat}}, \bibinfo {author}
  {\bibfnamefont {L.-J.}\ \bibnamefont {Wang}}, \bibinfo {author}
  {\bibfnamefont {N.}~\bibnamefont {Gregersen}}, \bibinfo {author}
  {\bibfnamefont {Y.-H.}\ \bibnamefont {Huo}}, \bibinfo {author} {\bibfnamefont
  {Q.}~\bibnamefont {Dai}}, \bibinfo {author} {\bibfnamefont {S.}~\bibnamefont
  {Yu}}, \bibinfo {author} {\bibfnamefont {S.}~\bibnamefont {H{\"o}fling}},
  \bibinfo {author} {\bibfnamefont {C.-Y.}\ \bibnamefont {Lu}}, \ and\ \bibinfo
  {author} {\bibfnamefont {J.-W.}\ \bibnamefont {Pan}},\ }\href {\doibase
  10.1038/s41566-019-0494-3} {\bibfield  {journal} {\bibinfo  {journal} {Nature
  Photonics}\ ,\ \bibinfo {pages} {1}} (\bibinfo {year} {2019})}\BibitemShut
  {NoStop}%
\bibitem [{\citenamefont {Giesz}\ \emph {et~al.}(2016)\citenamefont {Giesz},
  \citenamefont {Somaschi}, \citenamefont {Hornecker}, \citenamefont {Grange},
  \citenamefont {Reznychenko}, \citenamefont {De~Santis}, \citenamefont
  {Demory}, \citenamefont {Gomez}, \citenamefont {Sagnes}, \citenamefont
  {Lema{\^\i}tre}, \citenamefont {Krebs}, \citenamefont {Lanzillotti-Kimura},
  \citenamefont {Lanco}, \citenamefont {Auffeves},\ and\ \citenamefont
  {Senellart}}]{giesz_coherent_2016}%
  \BibitemOpen
  \bibfield  {author} {\bibinfo {author} {\bibfnamefont {V.}~\bibnamefont
  {Giesz}}, \bibinfo {author} {\bibfnamefont {N.}~\bibnamefont {Somaschi}},
  \bibinfo {author} {\bibfnamefont {G.}~\bibnamefont {Hornecker}}, \bibinfo
  {author} {\bibfnamefont {T.}~\bibnamefont {Grange}}, \bibinfo {author}
  {\bibfnamefont {B.}~\bibnamefont {Reznychenko}}, \bibinfo {author}
  {\bibfnamefont {L.}~\bibnamefont {De~Santis}}, \bibinfo {author}
  {\bibfnamefont {J.}~\bibnamefont {Demory}}, \bibinfo {author} {\bibfnamefont
  {C.}~\bibnamefont {Gomez}}, \bibinfo {author} {\bibfnamefont
  {I.}~\bibnamefont {Sagnes}}, \bibinfo {author} {\bibfnamefont
  {A.}~\bibnamefont {Lema{\^\i}tre}}, \bibinfo {author} {\bibfnamefont
  {O.}~\bibnamefont {Krebs}}, \bibinfo {author} {\bibfnamefont {N.~D.}\
  \bibnamefont {Lanzillotti-Kimura}}, \bibinfo {author} {\bibfnamefont
  {L.}~\bibnamefont {Lanco}}, \bibinfo {author} {\bibfnamefont
  {A.}~\bibnamefont {Auffeves}}, \ and\ \bibinfo {author} {\bibfnamefont
  {P.}~\bibnamefont {Senellart}},\ }\href {\doibase 10.1038/ncomms11986}
  {\bibfield  {journal} {\bibinfo  {journal} {Nature Communications}\ }\textbf
  {\bibinfo {volume} {7}},\ \bibinfo {pages} {11986} (\bibinfo {year}
  {2016})}\BibitemShut {NoStop}%
\bibitem [{\citenamefont {Loredo}\ \emph {et~al.}(2016)\citenamefont {Loredo},
  \citenamefont {Zakaria}, \citenamefont {Somaschi}, \citenamefont {Anton},
  \citenamefont {Santis}, \citenamefont {Giesz}, \citenamefont {Grange},
  \citenamefont {Broome}, \citenamefont {Gazzano}, \citenamefont {Coppola},
  \citenamefont {Sagnes}, \citenamefont {Lemaitre}, \citenamefont {Auffeves},
  \citenamefont {Senellart}, \citenamefont {Almeida},\ and\ \citenamefont
  {White}}]{loredo_scalable_2016}%
  \BibitemOpen
  \bibfield  {author} {\bibinfo {author} {\bibfnamefont {J.~C.}\ \bibnamefont
  {Loredo}}, \bibinfo {author} {\bibfnamefont {N.~A.}\ \bibnamefont {Zakaria}},
  \bibinfo {author} {\bibfnamefont {N.}~\bibnamefont {Somaschi}}, \bibinfo
  {author} {\bibfnamefont {C.}~\bibnamefont {Anton}}, \bibinfo {author}
  {\bibfnamefont {L.~d.}\ \bibnamefont {Santis}}, \bibinfo {author}
  {\bibfnamefont {V.}~\bibnamefont {Giesz}}, \bibinfo {author} {\bibfnamefont
  {T.}~\bibnamefont {Grange}}, \bibinfo {author} {\bibfnamefont {M.~A.}\
  \bibnamefont {Broome}}, \bibinfo {author} {\bibfnamefont {O.}~\bibnamefont
  {Gazzano}}, \bibinfo {author} {\bibfnamefont {G.}~\bibnamefont {Coppola}},
  \bibinfo {author} {\bibfnamefont {I.}~\bibnamefont {Sagnes}}, \bibinfo
  {author} {\bibfnamefont {A.}~\bibnamefont {Lemaitre}}, \bibinfo {author}
  {\bibfnamefont {A.}~\bibnamefont {Auffeves}}, \bibinfo {author}
  {\bibfnamefont {P.}~\bibnamefont {Senellart}}, \bibinfo {author}
  {\bibfnamefont {M.~P.}\ \bibnamefont {Almeida}}, \ and\ \bibinfo {author}
  {\bibfnamefont {A.~G.}\ \bibnamefont {White}},\ }\href {\doibase
  10.1364/OPTICA.3.000433} {\bibfield  {journal} {\bibinfo  {journal} {Optica}\
  }\textbf {\bibinfo {volume} {3}},\ \bibinfo {pages} {433} (\bibinfo {year}
  {2016})}\BibitemShut {NoStop}%
\bibitem [{\citenamefont {Grange}\ \emph {et~al.}(2017)\citenamefont {Grange},
  \citenamefont {Somaschi}, \citenamefont {Ant{\'o}n}, \citenamefont
  {De~Santis}, \citenamefont {Coppola}, \citenamefont {Giesz}, \citenamefont
  {Lema{\^\i}tre}, \citenamefont {Sagnes}, \citenamefont {Auff{\`e}ves},\ and\
  \citenamefont {Senellart}}]{grange_reducing_2017}%
  \BibitemOpen
  \bibfield  {author} {\bibinfo {author} {\bibfnamefont {T.}~\bibnamefont
  {Grange}}, \bibinfo {author} {\bibfnamefont {N.}~\bibnamefont {Somaschi}},
  \bibinfo {author} {\bibfnamefont {C.}~\bibnamefont {Ant{\'o}n}}, \bibinfo
  {author} {\bibfnamefont {L.}~\bibnamefont {De~Santis}}, \bibinfo {author}
  {\bibfnamefont {G.}~\bibnamefont {Coppola}}, \bibinfo {author} {\bibfnamefont
  {V.}~\bibnamefont {Giesz}}, \bibinfo {author} {\bibfnamefont
  {A.}~\bibnamefont {Lema{\^\i}tre}}, \bibinfo {author} {\bibfnamefont
  {I.}~\bibnamefont {Sagnes}}, \bibinfo {author} {\bibfnamefont
  {A.}~\bibnamefont {Auff{\`e}ves}}, \ and\ \bibinfo {author} {\bibfnamefont
  {P.}~\bibnamefont {Senellart}},\ }\href {\doibase
  10.1103/PhysRevLett.118.253602} {\bibfield  {journal} {\bibinfo  {journal}
  {Phys. Rev. Lett.}\ }\textbf {\bibinfo {volume} {118}},\ \bibinfo {pages}
  {253602} (\bibinfo {year} {2017})}\BibitemShut {NoStop}%
\bibitem [{\citenamefont {Yoshie}\ \emph {et~al.}(2004)\citenamefont {Yoshie},
  \citenamefont {Scherer}, \citenamefont {Hendrickson}, \citenamefont
  {Khitrova}, \citenamefont {Gibbs}, \citenamefont {Rupper}, \citenamefont
  {Ell}, \citenamefont {Shchekin},\ and\ \citenamefont
  {Deppe}}]{yoshie_vacuum_2004}%
  \BibitemOpen
  \bibfield  {author} {\bibinfo {author} {\bibfnamefont {T.}~\bibnamefont
  {Yoshie}}, \bibinfo {author} {\bibfnamefont {A.}~\bibnamefont {Scherer}},
  \bibinfo {author} {\bibfnamefont {J.}~\bibnamefont {Hendrickson}}, \bibinfo
  {author} {\bibfnamefont {G.}~\bibnamefont {Khitrova}}, \bibinfo {author}
  {\bibfnamefont {H.~M.}\ \bibnamefont {Gibbs}}, \bibinfo {author}
  {\bibfnamefont {G.}~\bibnamefont {Rupper}}, \bibinfo {author} {\bibfnamefont
  {C.}~\bibnamefont {Ell}}, \bibinfo {author} {\bibfnamefont {O.~B.}\
  \bibnamefont {Shchekin}}, \ and\ \bibinfo {author} {\bibfnamefont {D.~G.}\
  \bibnamefont {Deppe}},\ }\href {\doibase 10.1038/nature03119} {\bibfield
  {journal} {\bibinfo  {journal} {Nature}\ }\textbf {\bibinfo {volume} {432}},\
  \bibinfo {pages} {200} (\bibinfo {year} {2004})}\BibitemShut {NoStop}%
\bibitem [{\citenamefont {Huber}\ \emph {et~al.}(2017)\citenamefont {Huber},
  \citenamefont {Reindl}, \citenamefont {Huo}, \citenamefont {Huang},
  \citenamefont {Wildmann}, \citenamefont {Schmidt}, \citenamefont {Rastelli},\
  and\ \citenamefont {Trotta}}]{huber_highly_2017}%
  \BibitemOpen
  \bibfield  {author} {\bibinfo {author} {\bibfnamefont {D.}~\bibnamefont
  {Huber}}, \bibinfo {author} {\bibfnamefont {M.}~\bibnamefont {Reindl}},
  \bibinfo {author} {\bibfnamefont {Y.}~\bibnamefont {Huo}}, \bibinfo {author}
  {\bibfnamefont {H.}~\bibnamefont {Huang}}, \bibinfo {author} {\bibfnamefont
  {J.~S.}\ \bibnamefont {Wildmann}}, \bibinfo {author} {\bibfnamefont {O.~G.}\
  \bibnamefont {Schmidt}}, \bibinfo {author} {\bibfnamefont {A.}~\bibnamefont
  {Rastelli}}, \ and\ \bibinfo {author} {\bibfnamefont {R.}~\bibnamefont
  {Trotta}},\ }\href {\doibase 10.1038/ncomms15506} {\bibfield  {journal}
  {\bibinfo  {journal} {Nature Communications}\ }\textbf {\bibinfo {volume}
  {8}},\ \bibinfo {pages} {15506} (\bibinfo {year} {2017})}\BibitemShut
  {NoStop}%
\bibitem [{\citenamefont {Sch{\"o}ll}\ \emph {et~al.}(2019)\citenamefont
  {Sch{\"o}ll}, \citenamefont {Hanschke}, \citenamefont {Schweickert},
  \citenamefont {Zeuner}, \citenamefont {Reindl}, \citenamefont {da~Silva},
  \citenamefont {Lettner}, \citenamefont {Trotta}, \citenamefont {Finley},
  \citenamefont {M{\"u}ller}, \citenamefont {Rastelli}, \citenamefont
  {Zwiller},\ and\ \citenamefont {J{\"o}ns}}]{scholl_resonance_2019}%
  \BibitemOpen
  \bibfield  {author} {\bibinfo {author} {\bibfnamefont {E.}~\bibnamefont
  {Sch{\"o}ll}}, \bibinfo {author} {\bibfnamefont {L.}~\bibnamefont
  {Hanschke}}, \bibinfo {author} {\bibfnamefont {L.}~\bibnamefont
  {Schweickert}}, \bibinfo {author} {\bibfnamefont {K.~D.}\ \bibnamefont
  {Zeuner}}, \bibinfo {author} {\bibfnamefont {M.}~\bibnamefont {Reindl}},
  \bibinfo {author} {\bibfnamefont {S.~F.~C.}\ \bibnamefont {da~Silva}},
  \bibinfo {author} {\bibfnamefont {T.}~\bibnamefont {Lettner}}, \bibinfo
  {author} {\bibfnamefont {R.}~\bibnamefont {Trotta}}, \bibinfo {author}
  {\bibfnamefont {J.~J.}\ \bibnamefont {Finley}}, \bibinfo {author}
  {\bibfnamefont {K.}~\bibnamefont {M{\"u}ller}}, \bibinfo {author}
  {\bibfnamefont {A.}~\bibnamefont {Rastelli}}, \bibinfo {author}
  {\bibfnamefont {V.}~\bibnamefont {Zwiller}}, \ and\ \bibinfo {author}
  {\bibfnamefont {K.~D.}\ \bibnamefont {J{\"o}ns}},\ }\href {\doibase
  10.1021/acs.nanolett.8b05132} {\bibfield  {journal} {\bibinfo  {journal}
  {Nano Letters}\ }\textbf {\bibinfo {volume} {19}},\ \bibinfo {pages} {2404}
  (\bibinfo {year} {2019})},\ \bibinfo {note} {arXiv: 1901.09721}\BibitemShut
  {NoStop}%
\bibitem [{\citenamefont {Juska}\ \emph {et~al.}(2013)\citenamefont {Juska},
  \citenamefont {Dimastrodonato}, \citenamefont {Mereni}, \citenamefont
  {Gocalinska},\ and\ \citenamefont {Pelucchi}}]{juska_towards_2013}%
  \BibitemOpen
  \bibfield  {author} {\bibinfo {author} {\bibfnamefont {G.}~\bibnamefont
  {Juska}}, \bibinfo {author} {\bibfnamefont {V.}~\bibnamefont
  {Dimastrodonato}}, \bibinfo {author} {\bibfnamefont {L.~O.}\ \bibnamefont
  {Mereni}}, \bibinfo {author} {\bibfnamefont {A.}~\bibnamefont {Gocalinska}},
  \ and\ \bibinfo {author} {\bibfnamefont {E.}~\bibnamefont {Pelucchi}},\
  }\href {\doibase 10.1038/nphoton.2013.128} {\bibfield  {journal} {\bibinfo
  {journal} {Nature Photonics}\ }\textbf {\bibinfo {volume} {7}},\ \bibinfo
  {pages} {527} (\bibinfo {year} {2013})}\BibitemShut {NoStop}%
\bibitem [{\citenamefont {Claudon}\ \emph {et~al.}(2010)\citenamefont
  {Claudon}, \citenamefont {Bleuse}, \citenamefont {Malik}, \citenamefont
  {Bazin}, \citenamefont {Jaffrennou}, \citenamefont {Gregersen}, \citenamefont
  {Sauvan}, \citenamefont {Lalanne},\ and\ \citenamefont
  {G{\'e}rard}}]{claudon_highly_2010}%
  \BibitemOpen
  \bibfield  {author} {\bibinfo {author} {\bibfnamefont {J.}~\bibnamefont
  {Claudon}}, \bibinfo {author} {\bibfnamefont {J.}~\bibnamefont {Bleuse}},
  \bibinfo {author} {\bibfnamefont {N.~S.}\ \bibnamefont {Malik}}, \bibinfo
  {author} {\bibfnamefont {M.}~\bibnamefont {Bazin}}, \bibinfo {author}
  {\bibfnamefont {P.}~\bibnamefont {Jaffrennou}}, \bibinfo {author}
  {\bibfnamefont {N.}~\bibnamefont {Gregersen}}, \bibinfo {author}
  {\bibfnamefont {C.}~\bibnamefont {Sauvan}}, \bibinfo {author} {\bibfnamefont
  {P.}~\bibnamefont {Lalanne}}, \ and\ \bibinfo {author} {\bibfnamefont
  {J.-M.}\ \bibnamefont {G{\'e}rard}},\ }\href {\doibase
  10.1038/nphoton.2009.287x} {\bibfield  {journal} {\bibinfo  {journal} {Nature
  Photonics}\ }\textbf {\bibinfo {volume} {4}},\ \bibinfo {pages} {174}
  (\bibinfo {year} {2010})}\BibitemShut {NoStop}%
\bibitem [{\citenamefont {Zadeh}\ \emph {et~al.}(2016)\citenamefont {Zadeh},
  \citenamefont {Elshaari}, \citenamefont {J{\"o}ns}, \citenamefont {Fognini},
  \citenamefont {Dalacu}, \citenamefont {Poole}, \citenamefont {Reimer},\ and\
  \citenamefont {Zwiller}}]{zadeh_deterministic_2016}%
  \BibitemOpen
  \bibfield  {author} {\bibinfo {author} {\bibfnamefont {I.~E.}\ \bibnamefont
  {Zadeh}}, \bibinfo {author} {\bibfnamefont {A.~W.}\ \bibnamefont {Elshaari}},
  \bibinfo {author} {\bibfnamefont {K.~D.}\ \bibnamefont {J{\"o}ns}}, \bibinfo
  {author} {\bibfnamefont {A.}~\bibnamefont {Fognini}}, \bibinfo {author}
  {\bibfnamefont {D.}~\bibnamefont {Dalacu}}, \bibinfo {author} {\bibfnamefont
  {P.~J.}\ \bibnamefont {Poole}}, \bibinfo {author} {\bibfnamefont {M.~E.}\
  \bibnamefont {Reimer}}, \ and\ \bibinfo {author} {\bibfnamefont
  {V.}~\bibnamefont {Zwiller}},\ }\href {\doibase 10.1021/acs.nanolett.5b04709}
  {\bibfield  {journal} {\bibinfo  {journal} {Nano Letters}\ }\textbf {\bibinfo
  {volume} {16}},\ \bibinfo {pages} {2289} (\bibinfo {year}
  {2016})}\BibitemShut {NoStop}%
\bibitem [{\citenamefont {Kirsanske}\ \emph {et~al.}(2017)\citenamefont
  {Kirsanske}, \citenamefont {Thyrrestrup}, \citenamefont {Daveau},
  \citenamefont {Dree{\ss}en}, \citenamefont {Pregnolato}, \citenamefont
  {Midolo}, \citenamefont {Tighineanu}, \citenamefont {Javadi}, \citenamefont
  {Stobbe}, \citenamefont {Schott}, \citenamefont {Ludwig}, \citenamefont
  {Wieck}, \citenamefont {Park}, \citenamefont {Song}, \citenamefont
  {Kuhlmann}, \citenamefont {S{\"o}llner}, \citenamefont {L{\"o}bl},
  \citenamefont {Warburton},\ and\ \citenamefont
  {Lodahl}}]{kirsanske_indistinguishable_2017}%
  \BibitemOpen
  \bibfield  {author} {\bibinfo {author} {\bibfnamefont {G.}~\bibnamefont
  {Kirsanske}}, \bibinfo {author} {\bibfnamefont {H.}~\bibnamefont
  {Thyrrestrup}}, \bibinfo {author} {\bibfnamefont {R.~S.}\ \bibnamefont
  {Daveau}}, \bibinfo {author} {\bibfnamefont {C.~L.}\ \bibnamefont
  {Dree{\ss}en}}, \bibinfo {author} {\bibfnamefont {T.}~\bibnamefont
  {Pregnolato}}, \bibinfo {author} {\bibfnamefont {L.}~\bibnamefont {Midolo}},
  \bibinfo {author} {\bibfnamefont {P.}~\bibnamefont {Tighineanu}}, \bibinfo
  {author} {\bibfnamefont {A.}~\bibnamefont {Javadi}}, \bibinfo {author}
  {\bibfnamefont {S.}~\bibnamefont {Stobbe}}, \bibinfo {author} {\bibfnamefont
  {R.}~\bibnamefont {Schott}}, \bibinfo {author} {\bibfnamefont
  {A.}~\bibnamefont {Ludwig}}, \bibinfo {author} {\bibfnamefont {A.~D.}\
  \bibnamefont {Wieck}}, \bibinfo {author} {\bibfnamefont {S.~I.}\ \bibnamefont
  {Park}}, \bibinfo {author} {\bibfnamefont {J.~D.}\ \bibnamefont {Song}},
  \bibinfo {author} {\bibfnamefont {A.~V.}\ \bibnamefont {Kuhlmann}}, \bibinfo
  {author} {\bibfnamefont {I.}~\bibnamefont {S{\"o}llner}}, \bibinfo {author}
  {\bibfnamefont {M.~C.}\ \bibnamefont {L{\"o}bl}}, \bibinfo {author}
  {\bibfnamefont {R.~J.}\ \bibnamefont {Warburton}}, \ and\ \bibinfo {author}
  {\bibfnamefont {P.}~\bibnamefont {Lodahl}},\ }\href {\doibase
  10.1103/PhysRevB.96.165306} {\bibfield  {journal} {\bibinfo  {journal} {Phys.
  Rev. B}\ }\textbf {\bibinfo {volume} {96}},\ \bibinfo {pages} {165306}
  (\bibinfo {year} {2017})}\BibitemShut {NoStop}%
\bibitem [{\citenamefont {Ding}\ \emph {et~al.}(2016)\citenamefont {Ding},
  \citenamefont {He}, \citenamefont {Duan}, \citenamefont {Gregersen},
  \citenamefont {Chen}, \citenamefont {Unsleber}, \citenamefont {Maier},
  \citenamefont {Schneider}, \citenamefont {Kamp}, \citenamefont {H{\"o}fling},
  \citenamefont {Lu},\ and\ \citenamefont {Pan}}]{ding_-demand_2016}%
  \BibitemOpen
  \bibfield  {author} {\bibinfo {author} {\bibfnamefont {X.}~\bibnamefont
  {Ding}}, \bibinfo {author} {\bibfnamefont {Y.}~\bibnamefont {He}}, \bibinfo
  {author} {\bibfnamefont {Z.-C.}\ \bibnamefont {Duan}}, \bibinfo {author}
  {\bibfnamefont {N.}~\bibnamefont {Gregersen}}, \bibinfo {author}
  {\bibfnamefont {M.-C.}\ \bibnamefont {Chen}}, \bibinfo {author}
  {\bibfnamefont {S.}~\bibnamefont {Unsleber}}, \bibinfo {author}
  {\bibfnamefont {S.}~\bibnamefont {Maier}}, \bibinfo {author} {\bibfnamefont
  {C.}~\bibnamefont {Schneider}}, \bibinfo {author} {\bibfnamefont
  {M.}~\bibnamefont {Kamp}}, \bibinfo {author} {\bibfnamefont {S.}~\bibnamefont
  {H{\"o}fling}}, \bibinfo {author} {\bibfnamefont {C.-Y.}\ \bibnamefont {Lu}},
  \ and\ \bibinfo {author} {\bibfnamefont {J.-W.}\ \bibnamefont {Pan}},\ }\href
  {\doibase 10.1103/PhysRevLett.116.020401} {\bibfield  {journal} {\bibinfo
  {journal} {Phys. Rev. Lett.}\ }\textbf {\bibinfo {volume} {116}},\ \bibinfo
  {pages} {020401} (\bibinfo {year} {2016})}\BibitemShut {NoStop}%
\bibitem [{\citenamefont {Liu}\ \emph {et~al.}(2018)\citenamefont {Liu},
  \citenamefont {Brash}, \citenamefont {O'Hara}, \citenamefont {Martins},
  \citenamefont {Phillips}, \citenamefont {Coles}, \citenamefont {Royall},
  \citenamefont {Clarke}, \citenamefont {Bentham}, \citenamefont {Prtljaga},
  \citenamefont {Itskevich}, \citenamefont {Wilson}, \citenamefont {Skolnick},\
  and\ \citenamefont {Fox}}]{liu_high_2018}%
  \BibitemOpen
  \bibfield  {author} {\bibinfo {author} {\bibfnamefont {F.}~\bibnamefont
  {Liu}}, \bibinfo {author} {\bibfnamefont {A.~J.}\ \bibnamefont {Brash}},
  \bibinfo {author} {\bibfnamefont {J.}~\bibnamefont {O'Hara}}, \bibinfo
  {author} {\bibfnamefont {L.~M. P.~P.}\ \bibnamefont {Martins}}, \bibinfo
  {author} {\bibfnamefont {C.~L.}\ \bibnamefont {Phillips}}, \bibinfo {author}
  {\bibfnamefont {R.~J.}\ \bibnamefont {Coles}}, \bibinfo {author}
  {\bibfnamefont {B.}~\bibnamefont {Royall}}, \bibinfo {author} {\bibfnamefont
  {E.}~\bibnamefont {Clarke}}, \bibinfo {author} {\bibfnamefont
  {C.}~\bibnamefont {Bentham}}, \bibinfo {author} {\bibfnamefont
  {N.}~\bibnamefont {Prtljaga}}, \bibinfo {author} {\bibfnamefont {I.~E.}\
  \bibnamefont {Itskevich}}, \bibinfo {author} {\bibfnamefont {L.~R.}\
  \bibnamefont {Wilson}}, \bibinfo {author} {\bibfnamefont {M.~S.}\
  \bibnamefont {Skolnick}}, \ and\ \bibinfo {author} {\bibfnamefont {A.~M.}\
  \bibnamefont {Fox}},\ }\href@noop {} {\bibfield  {journal} {\bibinfo
  {journal} {Nature Nanotechnology}\ }\textbf {\bibinfo {volume} {13}},\
  \bibinfo {pages} {835} (\bibinfo {year} {2018})}\BibitemShut {NoStop}%
\bibitem [{\citenamefont {Purcell}\ \emph {et~al.}(1946)\citenamefont
  {Purcell}, \citenamefont {Torrey},\ and\ \citenamefont
  {Pound}}]{purcell_resonance_1946}%
  \BibitemOpen
  \bibfield  {author} {\bibinfo {author} {\bibfnamefont {E.~M.}\ \bibnamefont
  {Purcell}}, \bibinfo {author} {\bibfnamefont {H.~C.}\ \bibnamefont {Torrey}},
  \ and\ \bibinfo {author} {\bibfnamefont {R.~V.}\ \bibnamefont {Pound}},\
  }\href {\doibase 10.1103/PhysRev.69.37} {\bibfield  {journal} {\bibinfo
  {journal} {Phys. Rev.}\ }\textbf {\bibinfo {volume} {69}},\ \bibinfo {pages}
  {37} (\bibinfo {year} {1946})}\BibitemShut {NoStop}%
\bibitem [{\citenamefont {Auff{\`e}ves-Garnier}\ \emph
  {et~al.}(2007)\citenamefont {Auff{\`e}ves-Garnier}, \citenamefont {Simon},
  \citenamefont {G{\'e}rard},\ and\ \citenamefont
  {Poizat}}]{auffeves_giant_2007}%
  \BibitemOpen
  \bibfield  {author} {\bibinfo {author} {\bibfnamefont {A.}~\bibnamefont
  {Auff{\`e}ves-Garnier}}, \bibinfo {author} {\bibfnamefont {C.}~\bibnamefont
  {Simon}}, \bibinfo {author} {\bibfnamefont {J.-M.}\ \bibnamefont
  {G{\'e}rard}}, \ and\ \bibinfo {author} {\bibfnamefont {J.-P.}\ \bibnamefont
  {Poizat}},\ }\href {\doibase 10.1103/PhysRevA.75.053823} {\bibfield
  {journal} {\bibinfo  {journal} {Phys. Rev. A}\ }\textbf {\bibinfo {volume}
  {75}},\ \bibinfo {pages} {053823} (\bibinfo {year} {2007})}\BibitemShut
  {NoStop}%
\bibitem [{\citenamefont {Grange}\ \emph {et~al.}(2015)\citenamefont {Grange},
  \citenamefont {Hornecker}, \citenamefont {Hunger}, \citenamefont {Poizat},
  \citenamefont {G{\'e}rard}, \citenamefont {Senellart},\ and\ \citenamefont
  {Auff{\`e}ves}}]{grange_cavity-funneled_2015}%
  \BibitemOpen
  \bibfield  {author} {\bibinfo {author} {\bibfnamefont {T.}~\bibnamefont
  {Grange}}, \bibinfo {author} {\bibfnamefont {G.}~\bibnamefont {Hornecker}},
  \bibinfo {author} {\bibfnamefont {D.}~\bibnamefont {Hunger}}, \bibinfo
  {author} {\bibfnamefont {J.-P.}\ \bibnamefont {Poizat}}, \bibinfo {author}
  {\bibfnamefont {J.-M.}\ \bibnamefont {G{\'e}rard}}, \bibinfo {author}
  {\bibfnamefont {P.}~\bibnamefont {Senellart}}, \ and\ \bibinfo {author}
  {\bibfnamefont {A.}~\bibnamefont {Auff{\`e}ves}},\ }\href {\doibase
  10.1103/PhysRevLett.114.193601} {\bibfield  {journal} {\bibinfo  {journal}
  {Phys. Rev. Lett.}\ }\textbf {\bibinfo {volume} {114}},\ \bibinfo {pages}
  {193601} (\bibinfo {year} {2015})}\BibitemShut {NoStop}%
\bibitem [{\citenamefont {Iles-Smith}\ \emph {et~al.}(2017)\citenamefont
  {Iles-Smith}, \citenamefont {McCutcheon}, \citenamefont {Nazir},\ and\
  \citenamefont {M{\o}rk}}]{iles-smith_phonon_2017}%
  \BibitemOpen
  \bibfield  {author} {\bibinfo {author} {\bibfnamefont {J.}~\bibnamefont
  {Iles-Smith}}, \bibinfo {author} {\bibfnamefont {D.~P.~S.}\ \bibnamefont
  {McCutcheon}}, \bibinfo {author} {\bibfnamefont {A.}~\bibnamefont {Nazir}}, \
  and\ \bibinfo {author} {\bibfnamefont {J.}~\bibnamefont {M{\o}rk}},\ }\href
  {\doibase 10.1038/nphoton.2017.101} {\bibfield  {journal} {\bibinfo
  {journal} {Nature Photonics}\ }\textbf {\bibinfo {volume} {11}},\ \bibinfo
  {pages} {521} (\bibinfo {year} {2017})}\BibitemShut {NoStop}%
\bibitem [{\citenamefont {Davanco}\ \emph {et~al.}(2017)\citenamefont
  {Davanco}, \citenamefont {Liu}, \citenamefont {Sapienza}, \citenamefont
  {Zhang}, \citenamefont {Cardoso}, \citenamefont {Verma}, \citenamefont
  {Mirin}, \citenamefont {Nam}, \citenamefont {Liu},\ and\ \citenamefont
  {Srinivasan}}]{davanco_heterogeneous_2017}%
  \BibitemOpen
  \bibfield  {author} {\bibinfo {author} {\bibfnamefont {M.}~\bibnamefont
  {Davanco}}, \bibinfo {author} {\bibfnamefont {J.}~\bibnamefont {Liu}},
  \bibinfo {author} {\bibfnamefont {L.}~\bibnamefont {Sapienza}}, \bibinfo
  {author} {\bibfnamefont {C.-Z.}\ \bibnamefont {Zhang}}, \bibinfo {author}
  {\bibfnamefont {J.~V. D.~M.}\ \bibnamefont {Cardoso}}, \bibinfo {author}
  {\bibfnamefont {V.}~\bibnamefont {Verma}}, \bibinfo {author} {\bibfnamefont
  {R.}~\bibnamefont {Mirin}}, \bibinfo {author} {\bibfnamefont {S.~W.}\
  \bibnamefont {Nam}}, \bibinfo {author} {\bibfnamefont {L.}~\bibnamefont
  {Liu}}, \ and\ \bibinfo {author} {\bibfnamefont {K.}~\bibnamefont
  {Srinivasan}},\ }\href {\doibase 10.1038/s41467-017-00987-6} {\bibfield
  {journal} {\bibinfo  {journal} {Nature Communications}\ }\textbf {\bibinfo
  {volume} {8}},\ \bibinfo {pages} {889} (\bibinfo {year} {2017})}\BibitemShut
  {NoStop}%
\bibitem [{\citenamefont {He}\ \emph {et~al.}(2017)\citenamefont {He},
  \citenamefont {Liu}, \citenamefont {Maier}, \citenamefont {Emmerling},
  \citenamefont {Gerhardt}, \citenamefont {Davan{\c c}o}, \citenamefont
  {Srinivasan}, \citenamefont {Schneider},\ and\ \citenamefont
  {H{\"o}fling}}]{he_deterministic_2017}%
  \BibitemOpen
  \bibfield  {author} {\bibinfo {author} {\bibfnamefont {Y.-M.}\ \bibnamefont
  {He}}, \bibinfo {author} {\bibfnamefont {J.}~\bibnamefont {Liu}}, \bibinfo
  {author} {\bibfnamefont {S.}~\bibnamefont {Maier}}, \bibinfo {author}
  {\bibfnamefont {M.}~\bibnamefont {Emmerling}}, \bibinfo {author}
  {\bibfnamefont {S.}~\bibnamefont {Gerhardt}}, \bibinfo {author}
  {\bibfnamefont {M.}~\bibnamefont {Davan{\c c}o}}, \bibinfo {author}
  {\bibfnamefont {K.}~\bibnamefont {Srinivasan}}, \bibinfo {author}
  {\bibfnamefont {C.}~\bibnamefont {Schneider}}, \ and\ \bibinfo {author}
  {\bibfnamefont {S.}~\bibnamefont {H{\"o}fling}},\ }\href {\doibase
  10.1364/OPTICA.4.000802} {\bibfield  {journal} {\bibinfo  {journal} {Optica}\
  }\textbf {\bibinfo {volume} {4}},\ \bibinfo {pages} {802} (\bibinfo {year}
  {2017})}\BibitemShut {NoStop}%
\bibitem [{\citenamefont {Schnauber}\ \emph {et~al.}(2019)\citenamefont
  {Schnauber}, \citenamefont {Singh}, \citenamefont {Schall}, \citenamefont
  {Park}, \citenamefont {Song}, \citenamefont {Rodt}, \citenamefont
  {Srinivasan}, \citenamefont {Reitzenstein},\ and\ \citenamefont
  {Davanco}}]{schnauber_indistinguishable_2019}%
  \BibitemOpen
  \bibfield  {author} {\bibinfo {author} {\bibfnamefont {P.}~\bibnamefont
  {Schnauber}}, \bibinfo {author} {\bibfnamefont {A.}~\bibnamefont {Singh}},
  \bibinfo {author} {\bibfnamefont {J.}~\bibnamefont {Schall}}, \bibinfo
  {author} {\bibfnamefont {S.~I.}\ \bibnamefont {Park}}, \bibinfo {author}
  {\bibfnamefont {J.~D.}\ \bibnamefont {Song}}, \bibinfo {author}
  {\bibfnamefont {S.}~\bibnamefont {Rodt}}, \bibinfo {author} {\bibfnamefont
  {K.}~\bibnamefont {Srinivasan}}, \bibinfo {author} {\bibfnamefont
  {S.}~\bibnamefont {Reitzenstein}}, \ and\ \bibinfo {author} {\bibfnamefont
  {M.}~\bibnamefont {Davanco}},\ }\href {\doibase 10.1021/acs.nanolett.9b02758}
  {\bibfield  {journal} {\bibinfo  {journal} {Nano Letters}\ ,\ \bibinfo
  {pages} {acs.nanolett.9b02758}} (\bibinfo {year} {2019})}\BibitemShut
  {NoStop}%
\bibitem [{\citenamefont {Liu}\ \emph {et~al.}(2019)\citenamefont {Liu},
  \citenamefont {Su}, \citenamefont {Wei}, \citenamefont {Yao}, \citenamefont
  {Covre~da Silva}, \citenamefont {Yu}, \citenamefont {Iles-Smith},
  \citenamefont {Srinivasan}, \citenamefont {Rastelli}, \citenamefont {Li},\
  and\ \citenamefont {Wang}}]{liu_solid-state_2019}%
  \BibitemOpen
  \bibfield  {author} {\bibinfo {author} {\bibfnamefont {J.}~\bibnamefont
  {Liu}}, \bibinfo {author} {\bibfnamefont {R.}~\bibnamefont {Su}}, \bibinfo
  {author} {\bibfnamefont {Y.}~\bibnamefont {Wei}}, \bibinfo {author}
  {\bibfnamefont {B.}~\bibnamefont {Yao}}, \bibinfo {author} {\bibfnamefont
  {S.}~\bibnamefont {Covre~da Silva}}, \bibinfo {author} {\bibfnamefont
  {Y.}~\bibnamefont {Yu}}, \bibinfo {author} {\bibfnamefont {J.}~\bibnamefont
  {Iles-Smith}}, \bibinfo {author} {\bibfnamefont {K.}~\bibnamefont
  {Srinivasan}}, \bibinfo {author} {\bibfnamefont {A.}~\bibnamefont
  {Rastelli}}, \bibinfo {author} {\bibfnamefont {J.}~\bibnamefont {Li}}, \ and\
  \bibinfo {author} {\bibfnamefont {X.}~\bibnamefont {Wang}},\ }\href {\doibase
  10.1038/s41565-019-0435-9} {\bibfield  {journal} {\bibinfo  {journal} {Nature
  Nanotechnology}\ }\textbf {\bibinfo {volume} {14}},\ \bibinfo {pages} {1}
  (\bibinfo {year} {2019})}\BibitemShut {NoStop}%
\bibitem [{\citenamefont {Dousse}\ \emph {et~al.}(2008)\citenamefont {Dousse},
  \citenamefont {Lanco}, \citenamefont {Suffczynski}, \citenamefont {Semenova},
  \citenamefont {Miard}, \citenamefont {Lema{\^\i}tre}, \citenamefont {Sagnes},
  \citenamefont {Roblin}, \citenamefont {Bloch},\ and\ \citenamefont
  {Senellart}}]{dousse_controlled_2008}%
  \BibitemOpen
  \bibfield  {author} {\bibinfo {author} {\bibfnamefont {A.}~\bibnamefont
  {Dousse}}, \bibinfo {author} {\bibfnamefont {L.}~\bibnamefont {Lanco}},
  \bibinfo {author} {\bibfnamefont {J.}~\bibnamefont {Suffczynski}}, \bibinfo
  {author} {\bibfnamefont {E.}~\bibnamefont {Semenova}}, \bibinfo {author}
  {\bibfnamefont {A.}~\bibnamefont {Miard}}, \bibinfo {author} {\bibfnamefont
  {A.}~\bibnamefont {Lema{\^\i}tre}}, \bibinfo {author} {\bibfnamefont
  {I.}~\bibnamefont {Sagnes}}, \bibinfo {author} {\bibfnamefont
  {C.}~\bibnamefont {Roblin}}, \bibinfo {author} {\bibfnamefont
  {J.}~\bibnamefont {Bloch}}, \ and\ \bibinfo {author} {\bibfnamefont
  {P.}~\bibnamefont {Senellart}},\ }\href {\doibase
  10.1103/PhysRevLett.101.267404} {\bibfield  {journal} {\bibinfo  {journal}
  {Phys. Rev. Lett.}\ }\textbf {\bibinfo {volume} {101}},\ \bibinfo {pages}
  {267404} (\bibinfo {year} {2008})}\BibitemShut {NoStop}%
\bibitem [{\citenamefont {Ardelt}\ \emph {et~al.}(2015)\citenamefont {Ardelt},
  \citenamefont {Simmet}, \citenamefont {Müller}, \citenamefont {Dory},
  \citenamefont {Fischer}, \citenamefont {Bechtold}, \citenamefont {Kleinkauf},
  \citenamefont {Riedl},\ and\ \citenamefont
  {Finley}}]{ardelt_controlled_2015}%
  \BibitemOpen
  \bibfield  {author} {\bibinfo {author} {\bibfnamefont {P.-L.}\ \bibnamefont
  {Ardelt}}, \bibinfo {author} {\bibfnamefont {T.}~\bibnamefont {Simmet}},
  \bibinfo {author} {\bibfnamefont {K.}~\bibnamefont {Müller}}, \bibinfo
  {author} {\bibfnamefont {C.}~\bibnamefont {Dory}}, \bibinfo {author}
  {\bibfnamefont {K.}~\bibnamefont {Fischer}}, \bibinfo {author} {\bibfnamefont
  {A.}~\bibnamefont {Bechtold}}, \bibinfo {author} {\bibfnamefont
  {A.}~\bibnamefont {Kleinkauf}}, \bibinfo {author} {\bibfnamefont
  {H.}~\bibnamefont {Riedl}}, \ and\ \bibinfo {author} {\bibfnamefont
  {J.}~\bibnamefont {Finley}},\ }\href {\doibase 10.1103/PhysRevB.92.115306}
  {\bibfield  {journal} {\bibinfo  {journal} {Physical Review B}\ }\textbf
  {\bibinfo {volume} {92}} (\bibinfo {year} {2015}),\
  10.1103/PhysRevB.92.115306}\BibitemShut {NoStop}%
\bibitem [{\citenamefont {Hilaire}\ \emph {et~al.}(2019)\citenamefont
  {Hilaire}, \citenamefont {Millet}, \citenamefont {Loredo}, \citenamefont
  {Ant{\'o}n}, \citenamefont {Harouri}, \citenamefont {Lema{\^\i}tre},
  \citenamefont {Sagnes}, \citenamefont {Krebs}, \citenamefont {Senellart},\
  and\ \citenamefont {Lanco}}]{hilaire_2019}%
  \BibitemOpen
  \bibfield  {author} {\bibinfo {author} {\bibfnamefont {P.}~\bibnamefont
  {Hilaire}}, \bibinfo {author} {\bibfnamefont {C.}~\bibnamefont {Millet}},
  \bibinfo {author} {\bibfnamefont {J.~C.}\ \bibnamefont {Loredo}}, \bibinfo
  {author} {\bibfnamefont {C.}~\bibnamefont {Ant{\'o}n}}, \bibinfo {author}
  {\bibfnamefont {A.}~\bibnamefont {Harouri}}, \bibinfo {author} {\bibfnamefont
  {A.}~\bibnamefont {Lema{\^\i}tre}}, \bibinfo {author} {\bibfnamefont
  {I.}~\bibnamefont {Sagnes}}, \bibinfo {author} {\bibfnamefont {N.~S.~O.}\
  \bibnamefont {Krebs}}, \bibinfo {author} {\bibfnamefont {P.}~\bibnamefont
  {Senellart}}, \ and\ \bibinfo {author} {\bibfnamefont {L.}~\bibnamefont
  {Lanco}},\ }\href {http://arxiv.org/abs/1909.02440} {\bibfield  {journal}
  {\bibinfo  {journal} {arXiv:1909.02440 [cond-mat, physics:quant-ph]}\ }
  (\bibinfo {year} {2019})},\ \bibinfo {note} {arXiv: 1909.02440}\BibitemShut
  {NoStop}%
\bibitem [{\citenamefont {Nowak}\ \emph {et~al.}(2014)\citenamefont {Nowak},
  \citenamefont {Portalupi}, \citenamefont {Giesz}, \citenamefont {Gazzano},
  \citenamefont {Dal~Savio}, \citenamefont {Braun}, \citenamefont {Karrai},
  \citenamefont {Arnold}, \citenamefont {Lanco}, \citenamefont {Sagnes},
  \citenamefont {Lema{\^\i}tre},\ and\ \citenamefont
  {Senellart}}]{nowak_deterministic_2014}%
  \BibitemOpen
  \bibfield  {author} {\bibinfo {author} {\bibfnamefont {A.~K.}\ \bibnamefont
  {Nowak}}, \bibinfo {author} {\bibfnamefont {S.~L.}\ \bibnamefont
  {Portalupi}}, \bibinfo {author} {\bibfnamefont {V.}~\bibnamefont {Giesz}},
  \bibinfo {author} {\bibfnamefont {O.}~\bibnamefont {Gazzano}}, \bibinfo
  {author} {\bibfnamefont {C.}~\bibnamefont {Dal~Savio}}, \bibinfo {author}
  {\bibfnamefont {P.-F.}\ \bibnamefont {Braun}}, \bibinfo {author}
  {\bibfnamefont {K.}~\bibnamefont {Karrai}}, \bibinfo {author} {\bibfnamefont
  {C.}~\bibnamefont {Arnold}}, \bibinfo {author} {\bibfnamefont
  {L.}~\bibnamefont {Lanco}}, \bibinfo {author} {\bibfnamefont
  {I.}~\bibnamefont {Sagnes}}, \bibinfo {author} {\bibfnamefont
  {A.}~\bibnamefont {Lema{\^\i}tre}}, \ and\ \bibinfo {author} {\bibfnamefont
  {P.}~\bibnamefont {Senellart}},\ }\href {\doibase 10.1038/ncomms4240}
  {\bibfield  {journal} {\bibinfo  {journal} {Nature Communications}\ }\textbf
  {\bibinfo {volume} {5}},\ \bibinfo {pages} {3240} (\bibinfo {year}
  {2014})}\BibitemShut {NoStop}%
\bibitem [{\citenamefont {Bennett}\ \emph {et~al.}(2010)\citenamefont
  {Bennett}, \citenamefont {Patel}, \citenamefont {Skiba-Szymanska},
  \citenamefont {Nicoll}, \citenamefont {Farrer}, \citenamefont {Ritchie},\
  and\ \citenamefont {Shields}}]{bennett_giant_2010}%
  \BibitemOpen
  \bibfield  {author} {\bibinfo {author} {\bibfnamefont {A.~J.}\ \bibnamefont
  {Bennett}}, \bibinfo {author} {\bibfnamefont {R.~B.}\ \bibnamefont {Patel}},
  \bibinfo {author} {\bibfnamefont {J.}~\bibnamefont {Skiba-Szymanska}},
  \bibinfo {author} {\bibfnamefont {C.~A.}\ \bibnamefont {Nicoll}}, \bibinfo
  {author} {\bibfnamefont {I.}~\bibnamefont {Farrer}}, \bibinfo {author}
  {\bibfnamefont {D.~A.}\ \bibnamefont {Ritchie}}, \ and\ \bibinfo {author}
  {\bibfnamefont {A.~J.}\ \bibnamefont {Shields}},\ }\href {\doibase
  10.1063/1.3460912} {\bibfield  {journal} {\bibinfo  {journal} {Applied
  Physics Letters}\ }\textbf {\bibinfo {volume} {97}},\ \bibinfo {pages}
  {031104} (\bibinfo {year} {2010})}\BibitemShut {NoStop}%
\bibitem [{\citenamefont {Monniello}\ \emph {et~al.}(2014)\citenamefont
  {Monniello}, \citenamefont {Reigue}, \citenamefont {Hostein}, \citenamefont
  {Lemaitre}, \citenamefont {Martinez}, \citenamefont {Grousson},\ and\
  \citenamefont {Voliotis}}]{monniello_indistinguishable_2014}%
  \BibitemOpen
  \bibfield  {author} {\bibinfo {author} {\bibfnamefont {L.}~\bibnamefont
  {Monniello}}, \bibinfo {author} {\bibfnamefont {A.}~\bibnamefont {Reigue}},
  \bibinfo {author} {\bibfnamefont {R.}~\bibnamefont {Hostein}}, \bibinfo
  {author} {\bibfnamefont {A.}~\bibnamefont {Lemaitre}}, \bibinfo {author}
  {\bibfnamefont {A.}~\bibnamefont {Martinez}}, \bibinfo {author}
  {\bibfnamefont {R.}~\bibnamefont {Grousson}}, \ and\ \bibinfo {author}
  {\bibfnamefont {V.}~\bibnamefont {Voliotis}},\ }\href {\doibase
  10.1103/PhysRevB.90.041303} {\bibfield  {journal} {\bibinfo  {journal} {Phys.
  Rev. B}\ }\textbf {\bibinfo {volume} {90}},\ \bibinfo {pages} {041303}
  (\bibinfo {year} {2014})}\BibitemShut {NoStop}%
\bibitem [{\citenamefont {He}\ \emph {et~al.}(2013)\citenamefont {He},
  \citenamefont {He}, \citenamefont {Wei}, \citenamefont {Wu}, \citenamefont
  {Atat{\"u}re}, \citenamefont {Schneider}, \citenamefont {H{\"o}fling},
  \citenamefont {Kamp}, \citenamefont {Lu},\ and\ \citenamefont
  {Pan}}]{he_-demand_2013}%
  \BibitemOpen
  \bibfield  {author} {\bibinfo {author} {\bibfnamefont {Y.-M.}\ \bibnamefont
  {He}}, \bibinfo {author} {\bibfnamefont {Y.}~\bibnamefont {He}}, \bibinfo
  {author} {\bibfnamefont {Y.-J.}\ \bibnamefont {Wei}}, \bibinfo {author}
  {\bibfnamefont {D.}~\bibnamefont {Wu}}, \bibinfo {author} {\bibfnamefont
  {M.}~\bibnamefont {Atat{\"u}re}}, \bibinfo {author} {\bibfnamefont
  {C.}~\bibnamefont {Schneider}}, \bibinfo {author} {\bibfnamefont
  {S.}~\bibnamefont {H{\"o}fling}}, \bibinfo {author} {\bibfnamefont
  {M.}~\bibnamefont {Kamp}}, \bibinfo {author} {\bibfnamefont {C.-Y.}\
  \bibnamefont {Lu}}, \ and\ \bibinfo {author} {\bibfnamefont {J.-W.}\
  \bibnamefont {Pan}},\ }\href {\doibase 10.1038/nnano.2012.262} {\bibfield
  {journal} {\bibinfo  {journal} {Nature Nanotechnology}\ }\textbf {\bibinfo
  {volume} {8}},\ \bibinfo {pages} {213} (\bibinfo {year} {2013})}\BibitemShut
  {NoStop}%
\bibitem [{\citenamefont {Hilaire}\ \emph {et~al.}(2018)\citenamefont
  {Hilaire}, \citenamefont {Ant{\'o}n}, \citenamefont {Kessler}, \citenamefont
  {Lema{\^\i}tre}, \citenamefont {Sagnes}, \citenamefont {Somaschi},
  \citenamefont {Senellart},\ and\ \citenamefont
  {Lanco}}]{hilaire_accurate_2018}%
  \BibitemOpen
  \bibfield  {author} {\bibinfo {author} {\bibfnamefont {P.}~\bibnamefont
  {Hilaire}}, \bibinfo {author} {\bibfnamefont {C.}~\bibnamefont {Ant{\'o}n}},
  \bibinfo {author} {\bibfnamefont {C.}~\bibnamefont {Kessler}}, \bibinfo
  {author} {\bibfnamefont {A.}~\bibnamefont {Lema{\^\i}tre}}, \bibinfo {author}
  {\bibfnamefont {I.}~\bibnamefont {Sagnes}}, \bibinfo {author} {\bibfnamefont
  {N.}~\bibnamefont {Somaschi}}, \bibinfo {author} {\bibfnamefont
  {P.}~\bibnamefont {Senellart}}, \ and\ \bibinfo {author} {\bibfnamefont
  {L.}~\bibnamefont {Lanco}},\ }\href {\doibase 10.1063/1.5026799} {\bibfield
  {journal} {\bibinfo  {journal} {Applied Physics Letters}\ }\textbf {\bibinfo
  {volume} {112}},\ \bibinfo {pages} {201101} (\bibinfo {year}
  {2018})}\BibitemShut {NoStop}%
\bibitem [{\citenamefont {Bayer}\ \emph {et~al.}(2002)\citenamefont {Bayer},
  \citenamefont {Ortner}, \citenamefont {Stern}, \citenamefont {Kuther},
  \citenamefont {Gorbunov}, \citenamefont {Forchel}, \citenamefont {Hawrylak},
  \citenamefont {Fafard}, \citenamefont {Hinzer}, \citenamefont {Reinecke},
  \citenamefont {Walck}, \citenamefont {Reithmaier}, \citenamefont {Klopf},\
  and\ \citenamefont {Sch{\"a}fer}}]{bayer_fine_2002}%
  \BibitemOpen
  \bibfield  {author} {\bibinfo {author} {\bibfnamefont {M.}~\bibnamefont
  {Bayer}}, \bibinfo {author} {\bibfnamefont {G.}~\bibnamefont {Ortner}},
  \bibinfo {author} {\bibfnamefont {O.}~\bibnamefont {Stern}}, \bibinfo
  {author} {\bibfnamefont {A.}~\bibnamefont {Kuther}}, \bibinfo {author}
  {\bibfnamefont {A.~A.}\ \bibnamefont {Gorbunov}}, \bibinfo {author}
  {\bibfnamefont {A.}~\bibnamefont {Forchel}}, \bibinfo {author} {\bibfnamefont
  {P.}~\bibnamefont {Hawrylak}}, \bibinfo {author} {\bibfnamefont
  {S.}~\bibnamefont {Fafard}}, \bibinfo {author} {\bibfnamefont
  {K.}~\bibnamefont {Hinzer}}, \bibinfo {author} {\bibfnamefont {T.~L.}\
  \bibnamefont {Reinecke}}, \bibinfo {author} {\bibfnamefont {S.~N.}\
  \bibnamefont {Walck}}, \bibinfo {author} {\bibfnamefont {J.~P.}\ \bibnamefont
  {Reithmaier}}, \bibinfo {author} {\bibfnamefont {F.}~\bibnamefont {Klopf}}, \
  and\ \bibinfo {author} {\bibfnamefont {F.}~\bibnamefont {Sch{\"a}fer}},\
  }\href {\doibase 10.1103/PhysRevB.65.195315} {\bibfield  {journal} {\bibinfo
  {journal} {Phys. Rev. B}\ }\textbf {\bibinfo {volume} {65}},\ \bibinfo
  {pages} {195315} (\bibinfo {year} {2002})}\BibitemShut {NoStop}%
\bibitem [{\citenamefont {Bertlmann}\ \emph {et~al.}(2006)\citenamefont
  {Bertlmann}, \citenamefont {Grimus},\ and\ \citenamefont
  {Hiesmayr}}]{bertlmann_open-quantum-system_2006}%
  \BibitemOpen
  \bibfield  {author} {\bibinfo {author} {\bibfnamefont {R.~A.}\ \bibnamefont
  {Bertlmann}}, \bibinfo {author} {\bibfnamefont {W.}~\bibnamefont {Grimus}}, \
  and\ \bibinfo {author} {\bibfnamefont {B.~C.}\ \bibnamefont {Hiesmayr}},\
  }\href {\doibase 10.1103/PhysRevA.73.054101} {\bibfield  {journal} {\bibinfo
  {journal} {Physical Review A}\ }\textbf {\bibinfo {volume} {73}},\ \bibinfo
  {pages} {054101} (\bibinfo {year} {2006})}\BibitemShut {NoStop}%
\bibitem [{\citenamefont {Stievater}\ \emph {et~al.}(2001)\citenamefont
  {Stievater}, \citenamefont {Li}, \citenamefont {Steel}, \citenamefont
  {Gammon}, \citenamefont {Katzer}, \citenamefont {Park}, \citenamefont
  {Piermarocchi},\ and\ \citenamefont {Sham}}]{Stievater2001RabiOO}%
  \BibitemOpen
  \bibfield  {author} {\bibinfo {author} {\bibfnamefont {T.~H.}\ \bibnamefont
  {Stievater}}, \bibinfo {author} {\bibfnamefont {X.}~\bibnamefont {Li}},
  \bibinfo {author} {\bibfnamefont {D.~G.}\ \bibnamefont {Steel}}, \bibinfo
  {author} {\bibfnamefont {D.}~\bibnamefont {Gammon}}, \bibinfo {author}
  {\bibfnamefont {D.~S.}\ \bibnamefont {Katzer}}, \bibinfo {author}
  {\bibfnamefont {D.}~\bibnamefont {Park}}, \bibinfo {author} {\bibfnamefont
  {C.}~\bibnamefont {Piermarocchi}}, \ and\ \bibinfo {author} {\bibfnamefont
  {L.~J.}\ \bibnamefont {Sham}},\ }\href@noop {} {\bibfield  {journal}
  {\bibinfo  {journal} {Physical review letters}\ }\textbf {\bibinfo {volume}
  {87 13}},\ \bibinfo {pages} {133603} (\bibinfo {year} {2001})}\BibitemShut
  {NoStop}%
\bibitem [{\citenamefont {Fischer}\ \emph {et~al.}(2017)\citenamefont
  {Fischer}, \citenamefont {Hanschke}, \citenamefont {Wierzbowski},
  \citenamefont {Simmet}, \citenamefont {Dory}, \citenamefont {Finley},
  \citenamefont {Vuckovic},\ and\ \citenamefont
  {M{\"u}ller}}]{fischer_signatures_2017}%
  \BibitemOpen
  \bibfield  {author} {\bibinfo {author} {\bibfnamefont {K.~A.}\ \bibnamefont
  {Fischer}}, \bibinfo {author} {\bibfnamefont {L.}~\bibnamefont {Hanschke}},
  \bibinfo {author} {\bibfnamefont {J.}~\bibnamefont {Wierzbowski}}, \bibinfo
  {author} {\bibfnamefont {T.}~\bibnamefont {Simmet}}, \bibinfo {author}
  {\bibfnamefont {C.}~\bibnamefont {Dory}}, \bibinfo {author} {\bibfnamefont
  {J.~J.}\ \bibnamefont {Finley}}, \bibinfo {author} {\bibfnamefont
  {J.}~\bibnamefont {Vuckovic}}, \ and\ \bibinfo {author} {\bibfnamefont
  {K.}~\bibnamefont {M{\"u}ller}},\ }\href {https://doi.org/10.1038/nphys4052}
  {\bibfield  {journal} {\bibinfo  {journal} {Nature Physics}\ }\textbf
  {\bibinfo {volume} {13}},\ \bibinfo {pages} {649} (\bibinfo {year}
  {2017})}\BibitemShut {NoStop}%
\bibitem [{\citenamefont {Loredo}\ \emph {et~al.}(2019)\citenamefont {Loredo},
  \citenamefont {Antón}, \citenamefont {Reznychenko}, \citenamefont {Hilaire},
  \citenamefont {Harouri}, \citenamefont {Millet}, \citenamefont {Ollivier},
  \citenamefont {Somaschi}, \citenamefont {De~Santis}, \citenamefont
  {Lemaître}, \citenamefont {Sagnes}, \citenamefont {Lanco}, \citenamefont
  {Auffèves}, \citenamefont {Krebs},\ and\ \citenamefont
  {Senellart}}]{loredo_generation_2019}%
  \BibitemOpen
  \bibfield  {author} {\bibinfo {author} {\bibfnamefont {J.~C.}\ \bibnamefont
  {Loredo}}, \bibinfo {author} {\bibfnamefont {C.}~\bibnamefont {Antón}},
  \bibinfo {author} {\bibfnamefont {B.}~\bibnamefont {Reznychenko}}, \bibinfo
  {author} {\bibfnamefont {P.}~\bibnamefont {Hilaire}}, \bibinfo {author}
  {\bibfnamefont {A.}~\bibnamefont {Harouri}}, \bibinfo {author} {\bibfnamefont
  {C.}~\bibnamefont {Millet}}, \bibinfo {author} {\bibfnamefont
  {H.}~\bibnamefont {Ollivier}}, \bibinfo {author} {\bibfnamefont
  {N.}~\bibnamefont {Somaschi}}, \bibinfo {author} {\bibfnamefont
  {L.}~\bibnamefont {De~Santis}}, \bibinfo {author} {\bibfnamefont
  {A.}~\bibnamefont {Lemaître}}, \bibinfo {author} {\bibfnamefont
  {I.}~\bibnamefont {Sagnes}}, \bibinfo {author} {\bibfnamefont
  {L.}~\bibnamefont {Lanco}}, \bibinfo {author} {\bibfnamefont
  {A.}~\bibnamefont {Auffèves}}, \bibinfo {author} {\bibfnamefont
  {O.}~\bibnamefont {Krebs}}, \ and\ \bibinfo {author} {\bibfnamefont
  {P.}~\bibnamefont {Senellart}},\ }\href {\doibase 10.1038/s41566-019-0506-3}
  {\bibfield  {journal} {\bibinfo  {journal} {Nature Photonics}\ } (\bibinfo
  {year} {2019}),\ 10.1038/s41566-019-0506-3}\BibitemShut {NoStop}%
\bibitem [{\citenamefont {Ollivier}\ \emph {et~al.}()\citenamefont {Ollivier}
  \emph {et~al.}}]{g2HOM}%
  \BibitemOpen
  \bibfield  {author} {\bibinfo {author} {\bibfnamefont {H.}~\bibnamefont
  {Ollivier}} \emph {et~al.},\ }\href@noop {} {\bibinfo  {journal} {in
  preparation}\ }\BibitemShut {NoStop}%
\bibitem [{\citenamefont {Senellart}\ \emph {et~al.}(2017)\citenamefont
  {Senellart}, \citenamefont {Solomon},\ and\ \citenamefont
  {White}}]{senellart_high-performance_2017}%
  \BibitemOpen
\bibfield  {journal} {  }\bibfield  {author} {\bibinfo {author} {\bibfnamefont
  {P.}~\bibnamefont {Senellart}}, \bibinfo {author} {\bibfnamefont
  {G.}~\bibnamefont {Solomon}}, \ and\ \bibinfo {author} {\bibfnamefont
  {A.}~\bibnamefont {White}},\ }\href {https://doi.org/10.1038/nnano.2017.218}
  {\bibfield  {journal} {\bibinfo  {journal} {Nature Nanotechnology}\ }\textbf
  {\bibinfo {volume} {12}},\ \bibinfo {pages} {1026} (\bibinfo {year}
  {2017})}\BibitemShut {NoStop}%
\bibitem [{\citenamefont {Trotta}\ \emph {et~al.}(2015)\citenamefont {Trotta},
  \citenamefont {Mart{\'\i}n-S{\'a}nchez}, \citenamefont {Daruka},
  \citenamefont {Ortix},\ and\ \citenamefont
  {Rastelli}}]{trotta_energy-tunable_2015}%
  \BibitemOpen
  \bibfield  {author} {\bibinfo {author} {\bibfnamefont {R.}~\bibnamefont
  {Trotta}}, \bibinfo {author} {\bibfnamefont {J.}~\bibnamefont
  {Mart{\'\i}n-S{\'a}nchez}}, \bibinfo {author} {\bibfnamefont
  {I.}~\bibnamefont {Daruka}}, \bibinfo {author} {\bibfnamefont
  {C.}~\bibnamefont {Ortix}}, \ and\ \bibinfo {author} {\bibfnamefont
  {A.}~\bibnamefont {Rastelli}},\ }\href {\doibase
  10.1103/PhysRevLett.114.150502} {\bibfield  {journal} {\bibinfo  {journal}
  {Phys. Rev. Lett.}\ }\textbf {\bibinfo {volume} {114}},\ \bibinfo {pages}
  {150502} (\bibinfo {year} {2015})}\BibitemShut {NoStop}%
\bibitem [{\citenamefont {Weihs}\ \emph {et~al.}(2012)\citenamefont {Weihs},
  \citenamefont {Jayakumar}, \citenamefont {Predojevic}, \citenamefont {Huber},
  \citenamefont {Kauten},\ and\ \citenamefont
  {Solomon}}]{weihs_deterministic_2012}%
  \BibitemOpen
  \bibfield  {author} {\bibinfo {author} {\bibfnamefont {G.}~\bibnamefont
  {Weihs}}, \bibinfo {author} {\bibfnamefont {H.}~\bibnamefont {Jayakumar}},
  \bibinfo {author} {\bibfnamefont {A.}~\bibnamefont {Predojevic}}, \bibinfo
  {author} {\bibfnamefont {T.}~\bibnamefont {Huber}}, \bibinfo {author}
  {\bibfnamefont {T.}~\bibnamefont {Kauten}}, \ and\ \bibinfo {author}
  {\bibfnamefont {G.}~\bibnamefont {Solomon}}\ }(\bibinfo {year} {2012})\ pp.\
  \bibinfo {pages} {674--675}\BibitemShut {NoStop}%
\bibitem [{\citenamefont {Jöns}\ \emph {et~al.}(2013)\citenamefont {Jöns},
  \citenamefont {Atkinson}, \citenamefont {Müller}, \citenamefont {Heldmaier},
  \citenamefont {Ulrich}, \citenamefont {Schmidt},\ and\ \citenamefont
  {Michler}}]{jons_triggered_2013}%
  \BibitemOpen
  \bibfield  {author} {\bibinfo {author} {\bibfnamefont {K.~D.}\ \bibnamefont
  {Jöns}}, \bibinfo {author} {\bibfnamefont {P.}~\bibnamefont {Atkinson}},
  \bibinfo {author} {\bibfnamefont {M.}~\bibnamefont {Müller}}, \bibinfo
  {author} {\bibfnamefont {M.}~\bibnamefont {Heldmaier}}, \bibinfo {author}
  {\bibfnamefont {S.~M.}\ \bibnamefont {Ulrich}}, \bibinfo {author}
  {\bibfnamefont {O.~G.}\ \bibnamefont {Schmidt}}, \ and\ \bibinfo {author}
  {\bibfnamefont {P.}~\bibnamefont {Michler}},\ }\href {\doibase
  10.1021/nl303668z} {\bibfield  {journal} {\bibinfo  {journal} {Nano Letters}\
  }\textbf {\bibinfo {volume} {13}},\ \bibinfo {pages} {126} (\bibinfo {year}
  {2013})}\BibitemShut {NoStop}%
\bibitem [{\citenamefont {Reimer}\ \emph {et~al.}(2016)\citenamefont {Reimer},
  \citenamefont {Bulgarini}, \citenamefont {Fognini}, \citenamefont {Heeres},
  \citenamefont {Witek}, \citenamefont {Versteegh}, \citenamefont {Rubino},
  \citenamefont {Braun}, \citenamefont {Kamp}, \citenamefont {H\"ofling},
  \citenamefont {Dalacu}, \citenamefont {Lapointe}, \citenamefont {Poole},\
  and\ \citenamefont {Zwiller}}]{PhysRevB.93.195316}%
  \BibitemOpen
  \bibfield  {author} {\bibinfo {author} {\bibfnamefont {M.~E.}\ \bibnamefont
  {Reimer}}, \bibinfo {author} {\bibfnamefont {G.}~\bibnamefont {Bulgarini}},
  \bibinfo {author} {\bibfnamefont {A.}~\bibnamefont {Fognini}}, \bibinfo
  {author} {\bibfnamefont {R.~W.}\ \bibnamefont {Heeres}}, \bibinfo {author}
  {\bibfnamefont {B.~J.}\ \bibnamefont {Witek}}, \bibinfo {author}
  {\bibfnamefont {M.~A.~M.}\ \bibnamefont {Versteegh}}, \bibinfo {author}
  {\bibfnamefont {A.}~\bibnamefont {Rubino}}, \bibinfo {author} {\bibfnamefont
  {T.}~\bibnamefont {Braun}}, \bibinfo {author} {\bibfnamefont
  {M.}~\bibnamefont {Kamp}}, \bibinfo {author} {\bibfnamefont {S.}~\bibnamefont
  {H\"ofling}}, \bibinfo {author} {\bibfnamefont {D.}~\bibnamefont {Dalacu}},
  \bibinfo {author} {\bibfnamefont {J.}~\bibnamefont {Lapointe}}, \bibinfo
  {author} {\bibfnamefont {P.~J.}\ \bibnamefont {Poole}}, \ and\ \bibinfo
  {author} {\bibfnamefont {V.}~\bibnamefont {Zwiller}},\ }\href {\doibase
  10.1103/PhysRevB.93.195316} {\bibfield  {journal} {\bibinfo  {journal} {Phys.
  Rev. B}\ }\textbf {\bibinfo {volume} {93}},\ \bibinfo {pages} {195316}
  (\bibinfo {year} {2016})}\BibitemShut {NoStop}%
\end{thebibliography}%

\end{document}